\documentclass[prd,amsmath,amssymb,nofootinbib]{revtex4} 
\usepackage{graphicx} 
\usepackage{amsmath}
\usepackage{amsfonts,amsbsy}
\usepackage{amssymb}
\usepackage{breqn}

\newcommand{\be}{\begin{equation}}
\newcommand{\ee}{\end{equation}}
\newcommand{\bea}{\begin{eqnarray}}
\newcommand{\eea}{\end{eqnarray}}

\def\lsim{\mathrel{\rlap{\lower4pt\hbox{\hskip1pt$\sim$}}
    \raise1pt\hbox{$<$}}}                
\def\gsim{\mathrel{\rlap{\lower4pt\hbox{\hskip1pt$\sim$}}
    \raise1pt\hbox{$>$}}}                

\begin{document}

\title{Double inclusive small-$x$ gluon production and their azimuthal
  correlations in a biased ensemble}

\author{Gary Kapilevich}
\email{gkapilevich@gradcenter.cuny.edu}
\affiliation{The Graduate School and University Center The City University
of New York 365 Fifth Avenue New York NY 10016 USA}

\begin{abstract}
  We consider double $gg\to g$ production in the presence of a
  bias on the unintegrated gluon distribution of the colliding hadrons
  or nuclei. Such bias could be due to the selection of configurations
  with a greater number of gluons or higher mean transverse momentum
  squared or, more generally, due to a modified spectral shape of the
  gluon distribution in the hadrons. Hence, we consider reweighted
  functional averages over the stochastic ensemble of small-$x$
  gluons. We evaluate explicitly the double inclusive gluon transverse
  momentum spectrum in high-energy collisions, and their azimuthal
  correlations, for a few simple examples of biases.
\end{abstract}

\maketitle

\section{Introduction}

Observables in particle collision experiments are often evaluated in
{\em biased} event ensembles.  For example, one may consider a class
of events with greater than average multiplicity of produced paticles,
or events with high transverse energy deposition in a particular
rapidity window, and so on. The modification of observables in such
event ensembles, as compared to the minimum bias ensemble, provides
insight into particle production in QCD. More generally, one could
construct event ensembles using a {\em functional bias}. For
example, one can define a biased event ensembles ``labelled'' by a
prescribed function $\zeta(\vec k,y)$~\footnote{The function
  $\zeta(\vec k)$ corresponds to $\frac{1}{2} N_c^2 A_\perp
  (1-\eta^{-1}(\vec k))$, in terms of the function $\eta(\vec k)$
  introduced in eqs.~(\ref{eq:eta_Xs}, \ref{eq:t(k)}) below; it is
  proportional to the transverse area of the collision and to the
  dimension of the adjoint representation.}, such that event $i$ is
assigned a weight
\be
w_i = \exp \left( \int \frac{\mathrm{d}^2k}{(2\pi)^2}{\mathrm{d}y}\,\,
\zeta(\vec k,y)\, \frac{\mathrm{d}N_i}{\mathrm{d}^2k \mathrm{d}y} \Big/
\frac{\mathrm{d}N_{\text{mb}}}{\mathrm{d}^2k \mathrm{d}y}\right)~.
\ee
Here, $\frac{dN_i}{\mathrm{d}^2k  \mathrm{d}y}$ refers to the transverse
momentum distribution of particles in event $i$, and
$\frac{dN_{\text{mb}}}{\mathrm{d}^2k  \mathrm{d}y}$ is the (unweighted) average
over all events. One would then study how weighted averages,
\be
\langle O\rangle = \frac{\sum_i w_i \, O_i}{\sum_i w_i}~,
\ee
of various observables depend on the bias constructed via $\zeta(\vec
k, y)$.  In this paper, we relate expectation values in biased
ensembles in high-energy collisions to reweighted averages over the
field configurations of small-$x$ gluons in the colliding
particles. Specifically, we shall be interested in the double
inclusive gluon distribution and their azimuthal correlations.

Observables in high-energy scattering in QCD are computed by
expressing them in terms of expectation values of various Wilson line
operators $O$; see, for example, ref.~\cite{Weigert:2005us}. The
expectation value $\langle O\rangle$ corresponds to a statistical
average~\cite{Kovner:2005pe} over the distribution of small-$x$ gluon
fields. Hence, the Wilson lines from which $O$ is constructed are
computed in the soft gluon field sourced by the valence color charge
density $\rho$, which is the large component of the light-cone color
current due to the partons with large light-cone momenta~\cite{MV}:
\begin{equation}
  -\nabla_\perp^2 \int dx^- \, A^{+a}(x^-,\vec x_\perp) \equiv
  -\nabla_\perp^2 A^{+a}(\vec x_\perp) = g\rho^a(\vec x_\perp)~.
\end{equation}
We have assumed that the fast hadron propagates in the positive
$z$-direction, and that $\rho(\vec x_\perp)$ is the source in covariant
gauge. We may also compute the average leading twist (covariant gauge)
gluon distribution itself, via
\begin{equation}
\left<g^2 \mathrm{tr}\, A^+(\vec k)\, A^+(-\vec k)\right> =
\int\mathcal{D}\rho \, W[\rho]\, \frac{g^4}{k^4}\, \mathrm{tr}\,
\rho(\vec k)\, \rho(-\vec k)~.
\label{eq:VEV_A+^2}
\end{equation}
The weight functional is assumed to be normalized to $\int
\mathcal{D}\rho\, W[\rho]=1$.

The constraint effective potential for
\begin{equation}
X(\vec{k})\equiv g^2\mathrm{tr} \, A^{+}(\vec{k}) A^{+}(-\vec{k})
\end{equation}  
is given by~\cite{Dumitru:2017cwt}
\be
e^{-V_{\text{eff}}[X]} =
\int\mathcal{D}\rho\,\, \delta\left(X(\vec{k})-\frac{g^4}{k^4}
\mathrm{tr}\, \rho(\vec{k}) \rho(-\vec{k})\right)\, W[\rho]~.
\ee
This integrates out fluctations of $\rho$ which do not affect the
covariant gauge gluon distribution.  The most likely gluon
distribution from eq.~(\ref{eq:VEV_A+^2}) can then be obtained (at
leading power in $N_c$) as the stationary point of the effective
potential:
\be
\frac{\delta V_{\text{eff}}[X]}{\delta X(\vec q)}=0 ~~~~\to ~~~~
X_s(\vec q)~.
\ee
Given an observable which is a functional of $X(\vec q)$, the ensemble
average now reads
\be
\left< O[X] \right> = \int\mathcal{D}X\, e^{-V_{\text{eff}}[X]}\,
O[X]~.
\label{eq:EnsembleAverage_X}
\ee
For a Gaussian color charge density weight functional $W[\rho]$, one has~\cite{Dumitru:2017cwt}
\begin{equation}
V_{\text{eff}}[X(\vec k)] =
\int\frac{\mathrm{d}^2k}{(2\pi)^2} \left[ \frac{k^4}{g^4\mu^2(k)}X(\vec k) -
  \frac{1}{2}A_{\perp} N_c^2 \log X(\vec k)\right]~,
  \label{eq:Veff_Gauss}
\end{equation}
and
\begin{equation}
X_s(k) = \frac{1}{2} N_c^2 A_{\perp} \frac{g^4\mu^2(k)}{k^4}~,
  \label{eq:Xs_Gauss}
\end{equation}
where $A_\perp$ denotes the transverse area over which the gluon
distribution has been integrated over. The function $\mu^2(k)$
parameterizes the Gaussian ensemble for the color charge density:
$W[\rho] \sim \exp[- \int \mathrm{d}^2k/(2\pi)^2~
  \rho^a(\vec k) \rho^a(-\vec k) / 2\mu^2(\vec k)]$. However, the
corresponding effective potential for $X(\vec k)$ is not quadratic but
of ``linear minus log'' form\footnote{By a field redefinition,
  $V_{\text{eff}}[X(\vec k)]$ can be rewritten as a Liouville
  potential for $\phi(\vec k) = \log X(\vec k)/X_s(k)$, see
  ref.~\cite{Dumitru:2017cwt}}.

To probe configurations away from the peak of the distribution, it is
standard in statistical physics to compute biased (or reweighted)
expectation values:
\begin{equation}
\left<\mathcal{O}\right>_b = \int\mathcal{D}\rho \, W[\rho]
\, b[\rho]\, \mathcal{O}[\rho]~.
\end{equation}
Just like $W[\rho]$, the bias $b[\rho]$ in general is supposed to be a
gauge invariant functional of the color charge density. Here, we
impose the bias directly on the gluon distribution $X(\vec k)$:
\bea
V_{\text{eff}}[X(\vec k)] &\to&  V_{\text{eff}}[X(\vec k)] - \log
b[X(\vec k)]~, \\
\int\mathcal{D}X\, e^{-V_{\text{eff}}[X(k)]}\, O[X]
&\to&  \int\mathcal{D}X\, e^{-V_{\text{eff}}[X(k)]}\, b[X]\, O[X]~.
\label{eq:X_avg}
\eea
In particular, we choose $b[X]$ so that the most likely gluon
distribution in the reweighted ensemble is shifted to
\begin{equation}
  X_{s,b}(\vec k) = \eta(\vec k)\, X_s(k)~,
  \label{eq:eta_Xs}
\end{equation}
where $\eta(\vec k)\ge0$ is some prescribed function of transverse
momentum\footnote{We do require that the saddle point is not shifted to a
regime where the approach we described is not applicable. For example,
$X_{s,b}(\vec k)$ should not be of higher order in the coupling than
$X_s(k)$.}.
Defining
\be \label{eq:b[X]}
b[X] \equiv
\exp \left( \int\frac{\mathrm{d}^2\vec{k}}{(2\pi)^2}\  t(\vec{k})\,
X(\vec{k})\right)~,
\ee
this is achieved via
\be \label{eq:t_dVdX}
t(\vec{q}) = (2\pi)^2 \, \left.
\frac{\delta V_{\text{eff}}[X]}{\delta X(\vec q)}\right|_{X(\vec q) \,
  =\, \eta(\vec q)\, X_s(q^2)}~.
\ee
In fact, $b[X]$ is nothing but the generating functional for the
moments of $X(\vec k)$,
\be
Z[t] = \int \mathcal{D}X\, e^{-V_{\text{eff}}[X]+\log b[X]}~~~~,~~~~
\frac{1}{Z[t]} \frac{\delta^n Z[t]}{\delta t(\vec k_1)\cdots \delta t(\vec
  k_n)}\Biggr|_{t\equiv 0} =
\left< X(\vec k_1)\cdots X(\vec k_n)\right>~,
\ee
while $\log b[X]$ is the cumulant generating functional. 

In principle, the gluon distribution function depends on both the
transverse momentum and the rapidity, $y$. It is straightforward to
generalize the above to rapidity dependent biases by writing $X(\vec
q, y) = \eta(\vec q, y)\, X_s(q,y)$, so that $t(\vec q, y)$ also
depends on rapidity via eq.~(\ref{eq:t_dVdX}). One could then, for example, reweight
towards rare evolution trajectories. However, in this
paper we only consider the MV model~\cite{MV} effective theory of
color charge density fluctuations, which does not exhibit a dependence
on $y$.

For the Gaussian action for $\rho$, from
eqs.~(\ref{eq:Veff_Gauss},~\ref{eq:Xs_Gauss}) and~(\ref{eq:t_dVdX}), we
have, explicitly,
\begin{equation} \label{eq:t(k)}
t(\vec k) = \left(1-\frac{1}{\eta(\vec k)}\right)\frac{k^4}{g^4\mu^2(k)}~.
\end{equation}
A particularly simple example for a gluon distribution in a biased
ensemble would be
\be
X(\vec k) = \eta(\vec k)\, X_s(k^2)~~~~~,~~~~~
\eta(k) = 1 + \eta_0 \, \Theta\left(k^2 - \Lambda^2\right)\,
\Theta\left(Q^2 - k^2\right)~.
\ee
This simply boosts the number of gluons with transverse momenta from
$\Lambda^2$ to $Q^2$ by the constant factor $1+\eta_0$ (one may also
interpret this as a boost of the transverse momentum of the gluons by
a factor of $(1+\eta_0)^{1/4}$). Other examples will be considered
below.

To any given ``distortion'' $\eta(\vec k)$, one can associate a
potential, $V[\eta(\vec k)\, X_s(k^2)]$. The greater this potential,
the smaller the weight of the function $X(\vec k) = \eta(\vec k)\,
X_s(k)$ in the ensemble average~(\ref{eq:EnsembleAverage_X}). Hence,
a stronger bias is required to make this the dominant gluon
distribution in the reweighted ensemble. Explicitly, the ``penalty
action'' for any given $\eta(\vec k)$ is
\bea
\Delta V_{\text{eff}}[\eta(\vec k)] &\equiv& V_{\text{eff}}[\eta(\vec k)]
- V_{\text{eff}}[\eta=1]\notag  \\
&=&
\frac{1}{2} N_c^2 A_\perp \int\frac{\mathrm{d}^2k}{(2\pi)^2}
\left[ \eta(\vec k) -1 - \log\eta(\vec k)\right]~.
\label{eq:Veff_eta}
\eea
Thus, the gluon distribution $X(\vec k) = \eta(\vec k)\,
X_s(k)$ occurs, in the unbiased ensemble, with a probability density of
$p[\eta] = \exp (- \Delta V_{\text{eff}}[\eta])$, relative to the saddle point (in the space of functions). Note that $\eta(\vec
k)$ must be such that $\Delta V_{\text{eff}}[\eta(\vec k)]$ is finite.
Otherwise, the gluon distribution $X(\vec k) = \eta(\vec k)\,
X_s(k)$ is not part of the ensemble.

In the hadron or nucleus, a given $\eta(\vec k)$ also corresponds to an excess gluon multiplicity of~\cite{Dumitru:2017cwt}
\be \label{eq:DNg}
\Delta N_g[\eta(\vec k)] = \int\frac{\mathrm{d}^2k}{(2\pi)^2}\,
k^2 X_s(k)\, \left[\eta(\vec k)-1\right]~.
\ee
Likewise, any $\eta(\vec k)$ can also be associated with, for example, an increased mean
transverse momentum (see definition of $\left<k_T^2\right>$ in
ref.~\cite{Baier:1996sk}, for example).
We note, however, that our approach allows us to compute expectation
values in an ensemble defined by a {\em functional} bias on the
gluon distribution $X(\vec k)$, rather than simply an ensemble defined with a bias on gluon
number, mean transverse momentum, etc.  

One may sample the gluon distributions in a biased ensemble in the form of
eq.~(\ref{eq:X_avg}) via a Metropolis algorithm. While these gluon
distributions are part of the original ensemble, the standard approach
of generating configurations without bias and then rejecting those
that do not meet a given criteria would be prohibitive. Importance sampling
with the action $V_{\text{eff}}[X] - \log b[X]$ strongly increases the
overlap with the desired target ensemble. We consider the following
three biases for illustration:
\begin{enumerate}
\item $N_g$ bias corresponding to
  \be
  \log b[X] = N_g[X] = \int\limits_{\Lambda}^{Q}
  \frac{\mathrm{d}^2k}{(2\pi)^2}\,  k^2 X(k)~.
  \ee
  We take $\Lambda=2$ and $Q=6$; the units may be taken as GeV,
  although the energy scale is arbitrary, since $b[X]$ is
  dimensionless.  Also, we choose $A_\perp = 10 \pi$ and $g^4\mu^2 =
  2$ in eq.~(\ref{eq:Veff_Gauss}). This bias does not impose a
  specific transverse momentum dependence on $\langle
  X(k)\rangle_b$. Rather, we let the Monte-Carlo determine the optimal
  spectral shape.

\item $E_T$ bias corresponding to
  \be
  \log b[X] = \frac{E_T[X]}{\Lambda}
  = \int\limits_{\Lambda}^{Q}
  \frac{\mathrm{d}^2k}{(2\pi)^2\, \Lambda}\,  k^3 \, X(k)~.
  \ee
  Once again, we do not impose a specific transverse
  momentum distribution on the gluons, and instead let the Monte-Carlo
  determine the optimal spectral shape.
\item $t[\eta]$ bias corresponding to
  \be
  \log b[X] = \int\limits_{\Lambda}^{Q}
  \frac{\mathrm{d}^2k}{(2\pi)^2}\  t(\vec{k})\, X(\vec{k})~,
  \ee
  with $t(\vec k) =
  \left(1-\eta^{-1}(k)\right)\frac{k^4}{g^4\mu^2}$ and the prescribed function
  $\eta(k) = \sqrt{k/\Lambda}$.
\end{enumerate}
In all cases the unbiased ensemble is taken to be the MV model with
constant $\mu^2$.

\begin{figure}[htb]
\includegraphics [scale=.6, trim={1cm 14.28cm 4.7cm 1cm},clip]{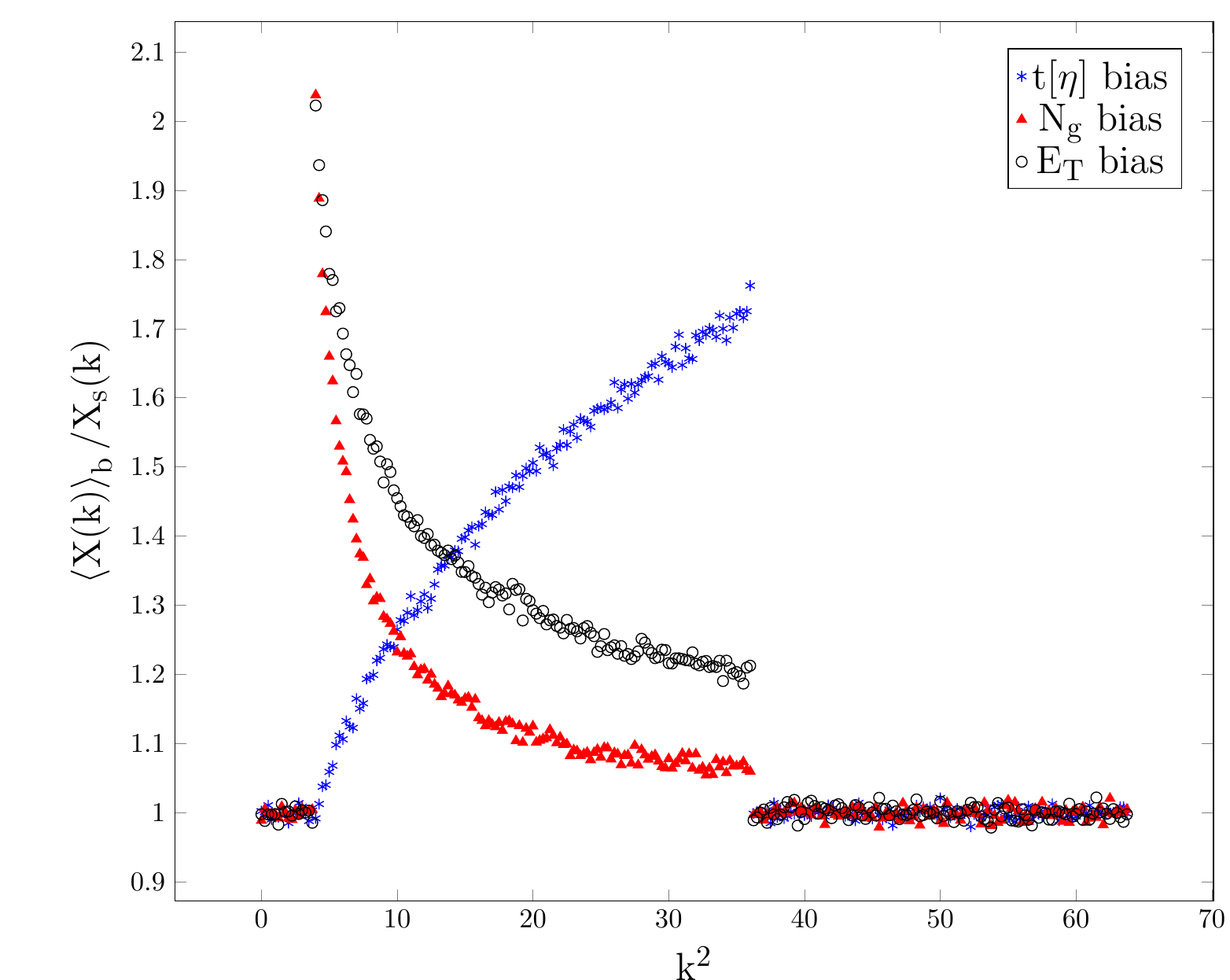}
\caption{Ratio of the gluon distribution in three different biased
  ensembles to that in the unbiased MV-model ensemble.}
\label{fig:Xmetropolis}
\end{figure}
Fig.~\ref{fig:Xmetropolis} shows the results. Not surprisingly, the
$N_g$ bias adds gluons mostly near $\Lambda$, since high-$k$ gluons
come with a greater penalty action. The analytic solution for this bias
is $\langle X(k)\rangle_b/X_s(k) = 1/[1-\frac{g^4\mu^2}{k^2}
  \Theta(Q^2-k^2)\, \Theta(k^2-\Lambda^2)]$. The $E_T$-bias produces a
harder spectrum of excess gluons, with $\langle X(k)\rangle_b/X_s(k) =
1/[1-\frac{g^4\mu^2}{k\,\Lambda} \Theta(Q^2-k^2)\,
  \Theta(k^2-\Lambda^2)]$. Lastly, the $t[\eta]$ bias multiplies the
gluon distribution between $\Lambda$ and $Q$ by the prescribed
function $\eta(k) = \sqrt{k/\Lambda}$.  \\

In a collision of two hadrons or nuclei, one is required to average
over the color charge distributions of both projectile and target,
\begin{equation}
\left<\mathcal{O}\right> =\int\mathcal{D}\rho_p\,
W[\rho_p]\int\mathcal{D}\rho_T \, W[\rho_T] \, \mathcal{O}[\rho_p,\rho_T]~.
\end{equation}
One may then bias either one or both of the ensembles as described above.

The single-inclusive gluon production cross section in a biased
ensemble has been computed previously in ref.~\cite{Dumitru:2018iko}.
The main purpose of the current paper is to illustrate the effect of a bias
on azimuthal angular correlations of two small-$x$, high-$p_T$ gluons
produced in a high-energy collision. We recompute the so-called
``glasma graphs'' for a biased gluon distribution different from its
expectation value $X_s(k)$ in the unbiased small-$x$ ensemble. These
diagrams for high-$p_T$ double gluon production have originally been
introduced in refs.~\cite{Dumitru:2008wn,Dusling:2009ni}. Their
applicability, and corrections to this approximation, have been
studied in
refs.~\cite{Kovchegov:2012nd,Lappi:2015vta,Altinoluk:2018hcu}.

The literature on azimuthal correlations of small-$x$ gluons is rather
extensive and we do not attempt to summarize it here. Instead, we
refer the reader to the review articles in
refs~\cite{Kovner:2012jm,Schlichting:2016sqo}. Our main focus now is
on effects due to a bias on the gluon distributions of the colliding
hadrons or nuclei, an issue which has rarely been addressed. A notable
exception is ref.~\cite{Dusling:2012iga}, where the authors assumed
that high multiplicity p+p and p+Pb events correspond to an enhanced
saturation scale, $Q_s(x_0)$, of the proton at the initial rapidity for
small-x evolution. Ref.~\cite{Mace:2018yvl} considered a constant
multiplicative rescaling of the color charge density in the proton to
discuss the multiplicity dependence of azimuthal moments (as defined
  in eq.~(\ref{eq:ang_avg}) below) in p+Pb
collisions. Ref.~\cite{Blok:2017pui} analyzed angular correlations in
a combinatoric model for multi-particle production with color
interference effects, and their dependence on
multiplicity. Ref.~\cite{Schenke:2019pmk}, finally, applied a
hydrodynamic model to look into the effect of final state interactions
on angular correlations, as a function of the particle multiplicity in
the event. Here, we perform the first analysis of ``glasma graphs''
in the presence of a {\em functional} bias on the gluon
distribution.

The remainder of this paper is organized as follows.  In
section~\ref{sec:General_eta}, we write the two gluon inclusive
distribution at high transverse momentum for general $\eta(\vec k)$.
In sec.~\ref{sec:ModelEnsembles}, we analyze several specific
momentum dependences to see how they affect the double gluon spectrum
and their angular correlations.

\section{Two gluon inclusive distribution in a biased ensemble}
\label{sec:General_eta}

The cross section for inclusive production of two small-$x$ gluons
with transverse momenta $p$, $q$ much greater than the saturation
scales of the projectile and target is given by so-called ``glasma
graphs''. These graphs correspond to a $k_T$-factorization
approximation in terms of unintegrated gluon distributions~\cite{kTfact},
\be
\Phi(k) = \frac{1}{A_\perp} \frac{k^2}{N_c^2-1}\, \left<
A^{+a}(\vec k)\, A^{+a}(-\vec k)\right> =
\frac{g^2\mu^2}{k^2} ~.
\ee
From now on, we consider a constant, $k$-independent $\mu^2$ for
simplicity. This amounts to the classical MV model~\cite{MV}
approximation where one neglects the anomalous dimension of the gluon
distribution. While it is possible, in principle, to generalize our
analysis to account for the anomalous dimension due to small-$x$
evolution, our current focus is on better understanding the effect of
a bias on the glasma graphs.

\begin{figure}[t]
\includegraphics[scale=.15]{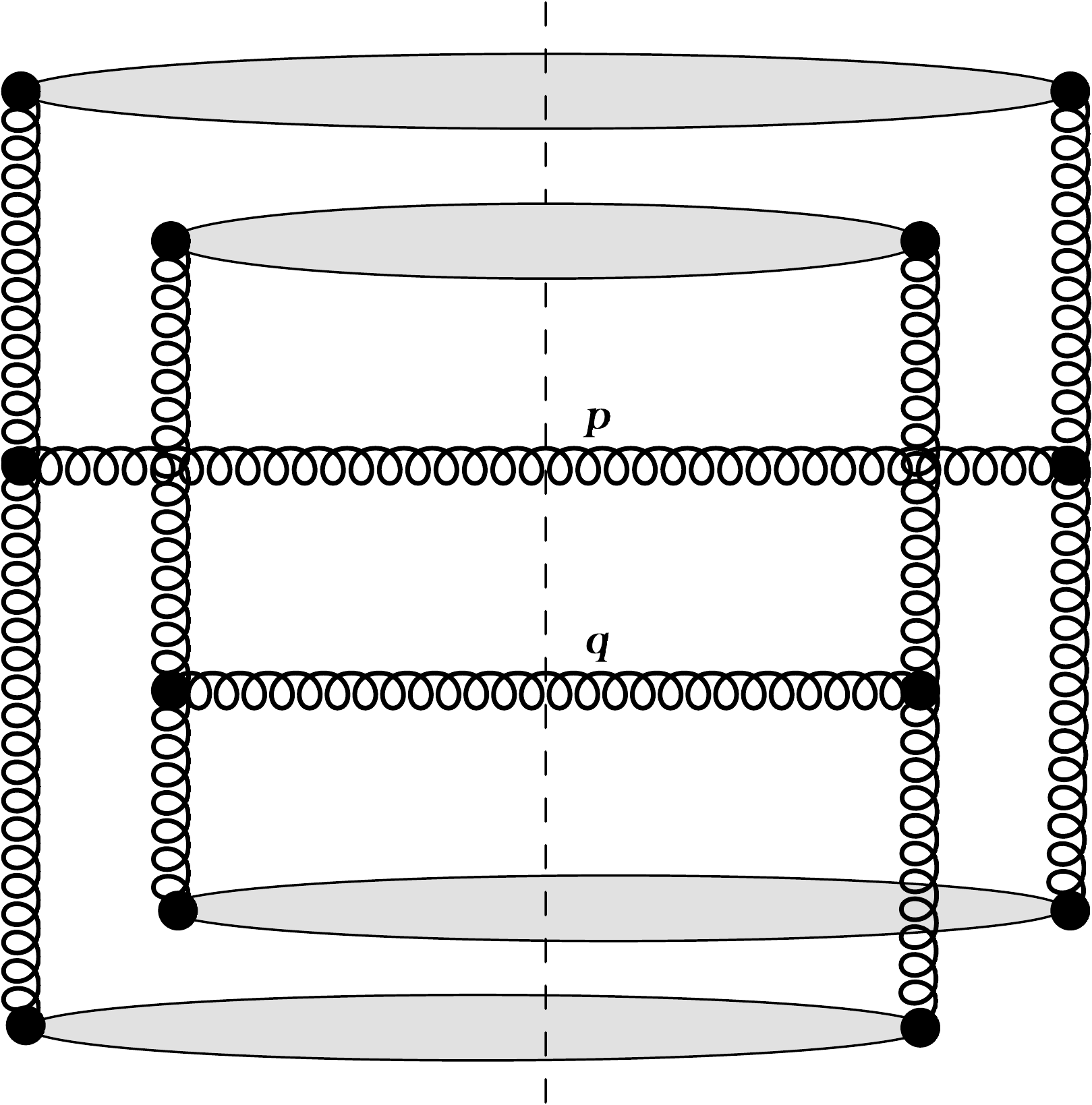}~~~~~~
\includegraphics[scale=.15]{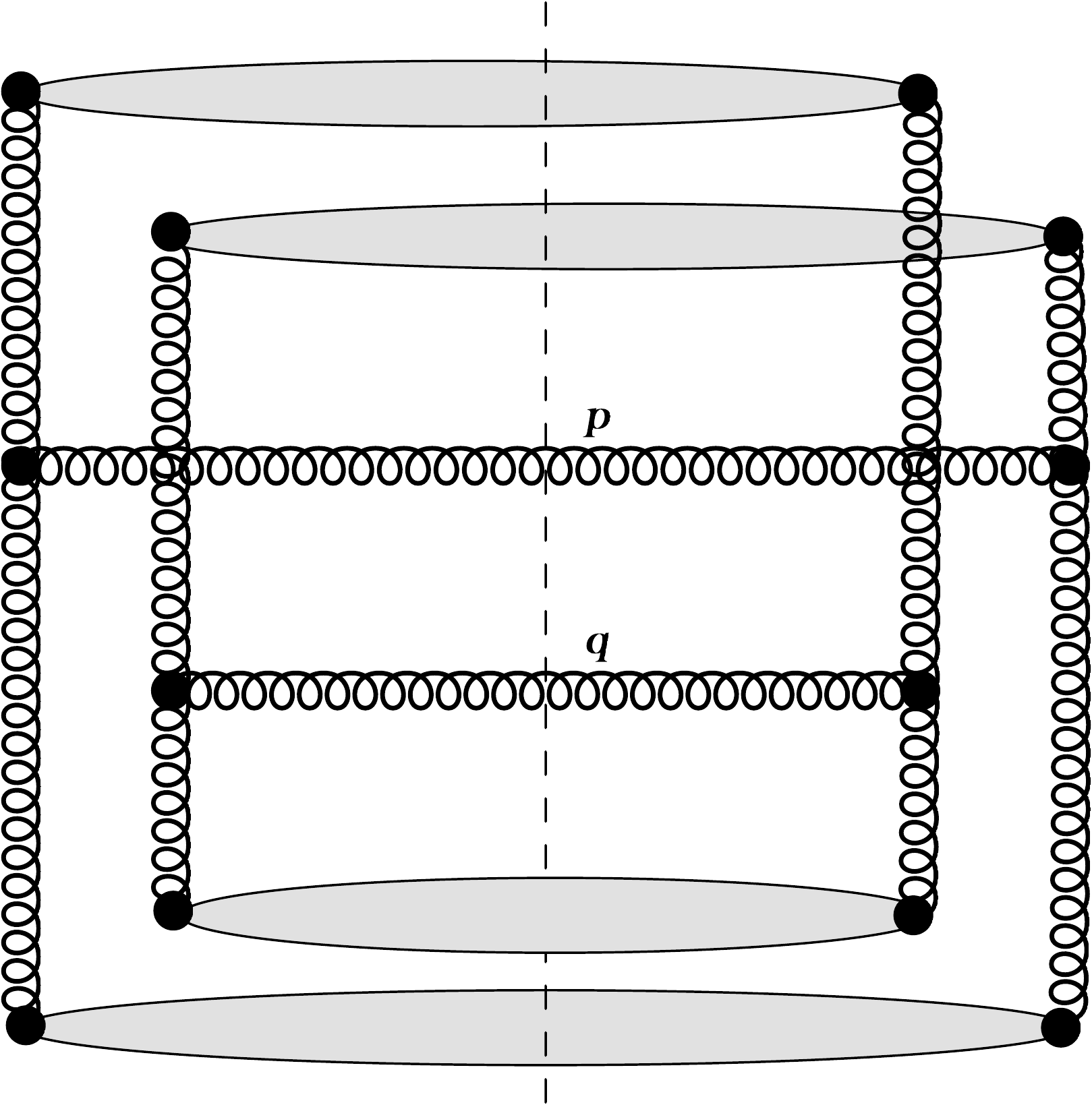}~~~~~~
\includegraphics[scale=.14]{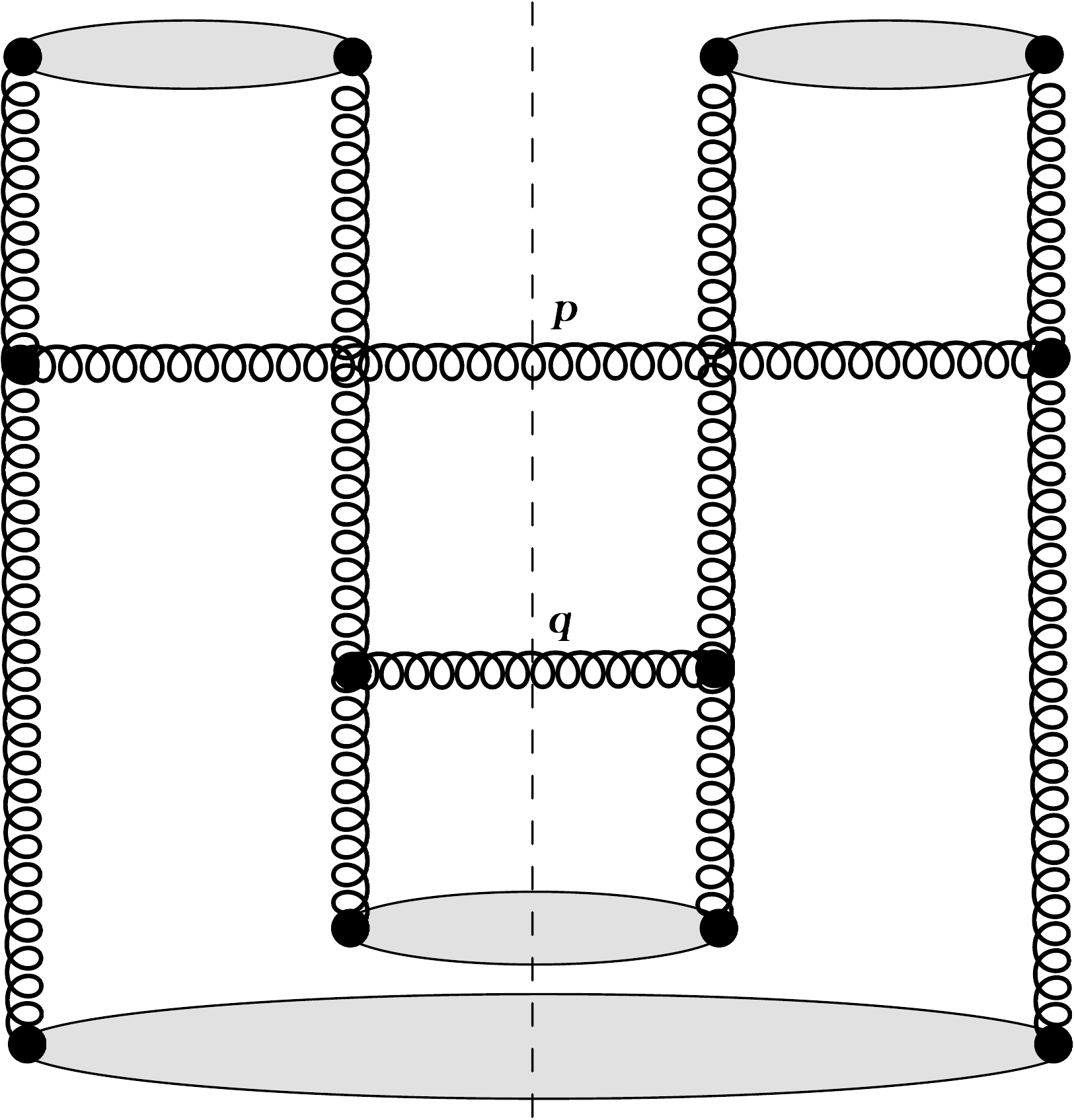}~~~~~~
\includegraphics[scale=.14]{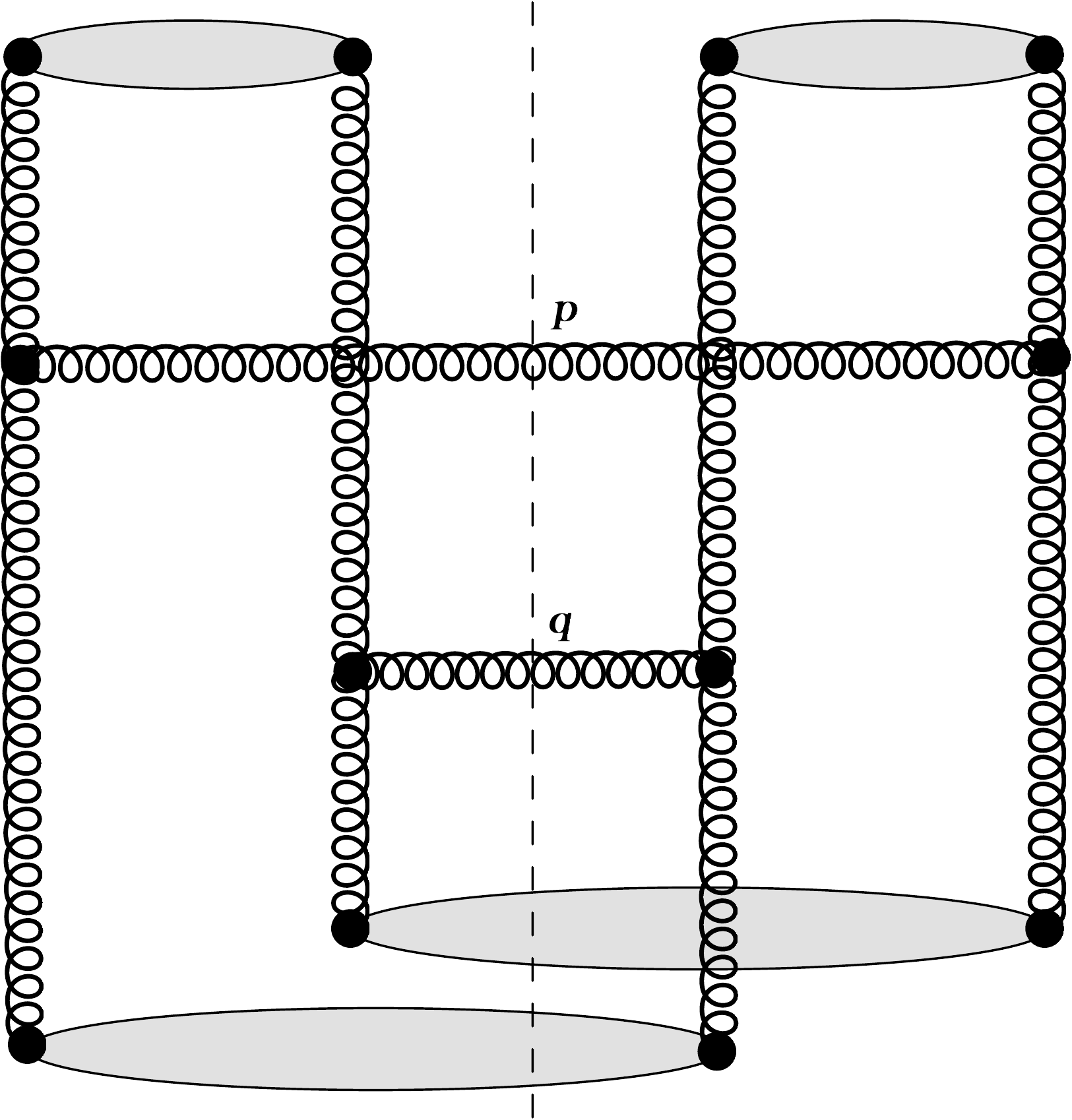}~~~~~~
\includegraphics[scale=.14]{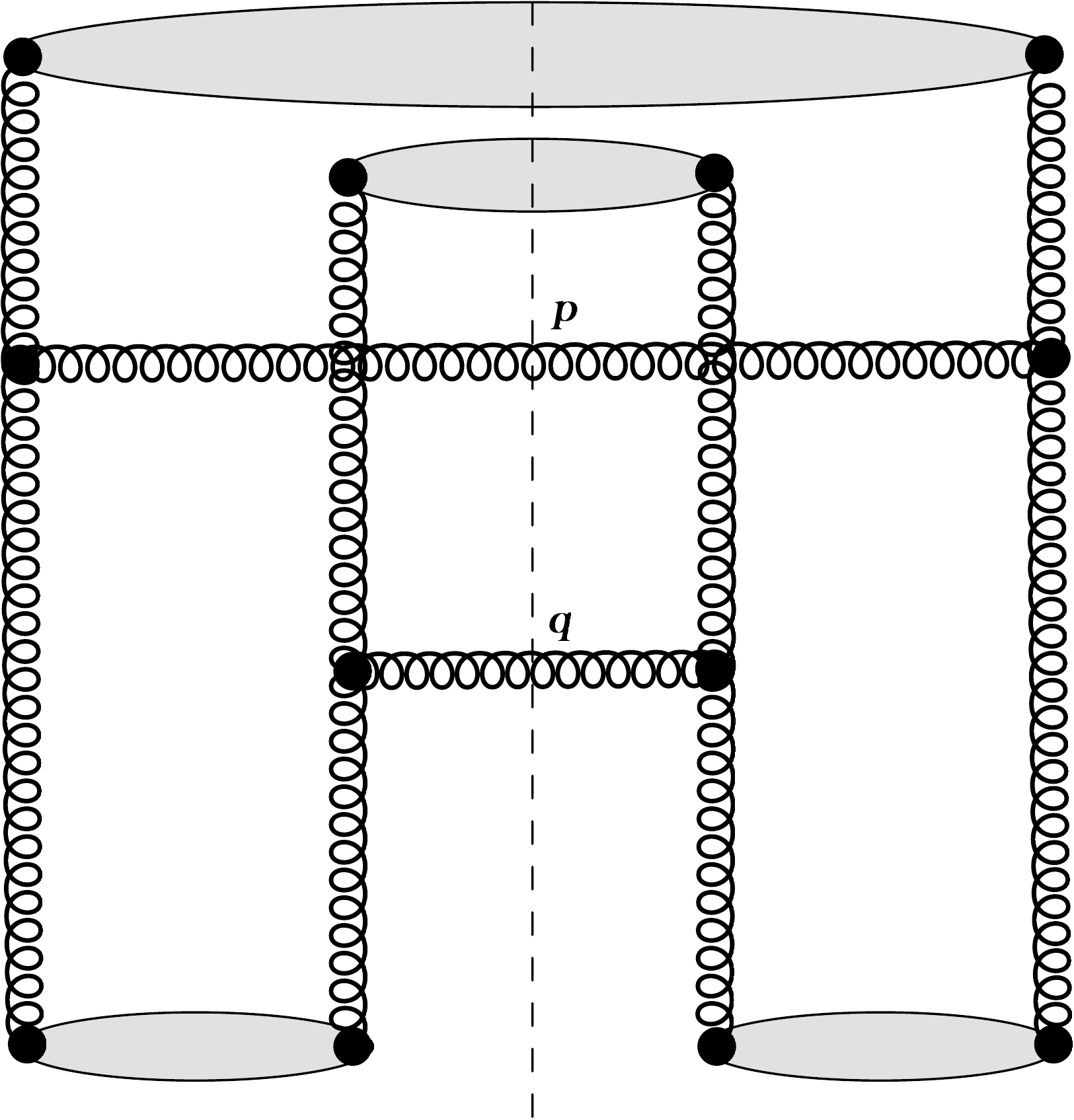}~~~~~~
\includegraphics[scale=.14]{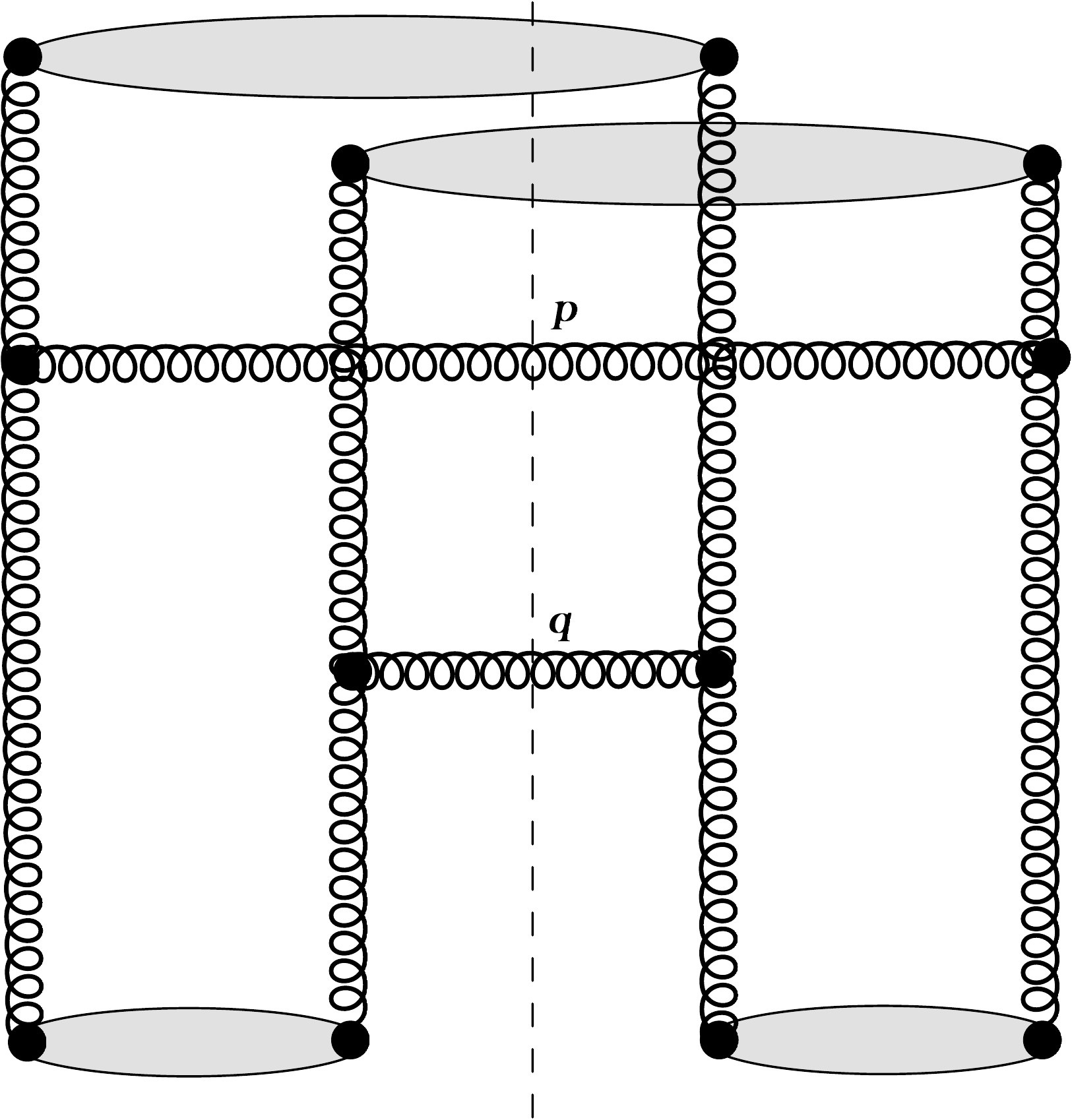}~~~~~~
\caption{Bose enhancement ($\mathcal{B}$-type) diagrams.}
\label{fig:B_diagrams}
\end{figure}

In a biased ensemble,
\be \Phi_b(\vec {k})=\frac{g^2\mu^2}{k^2}\, \eta(\vec{k})~.
\ee
Beyond the dilute limit, one needs to evaluate the correlator of two
eikonal Wilson lines in the reweighted ensemble, see
ref.~\cite{Dumitru:2018iko}. Here, we restrict our analysis to high transverse
momentum, where the approximation of a dilute projectile and target
should be applicable.
\begin{figure}[htb]
\includegraphics[scale=.2]{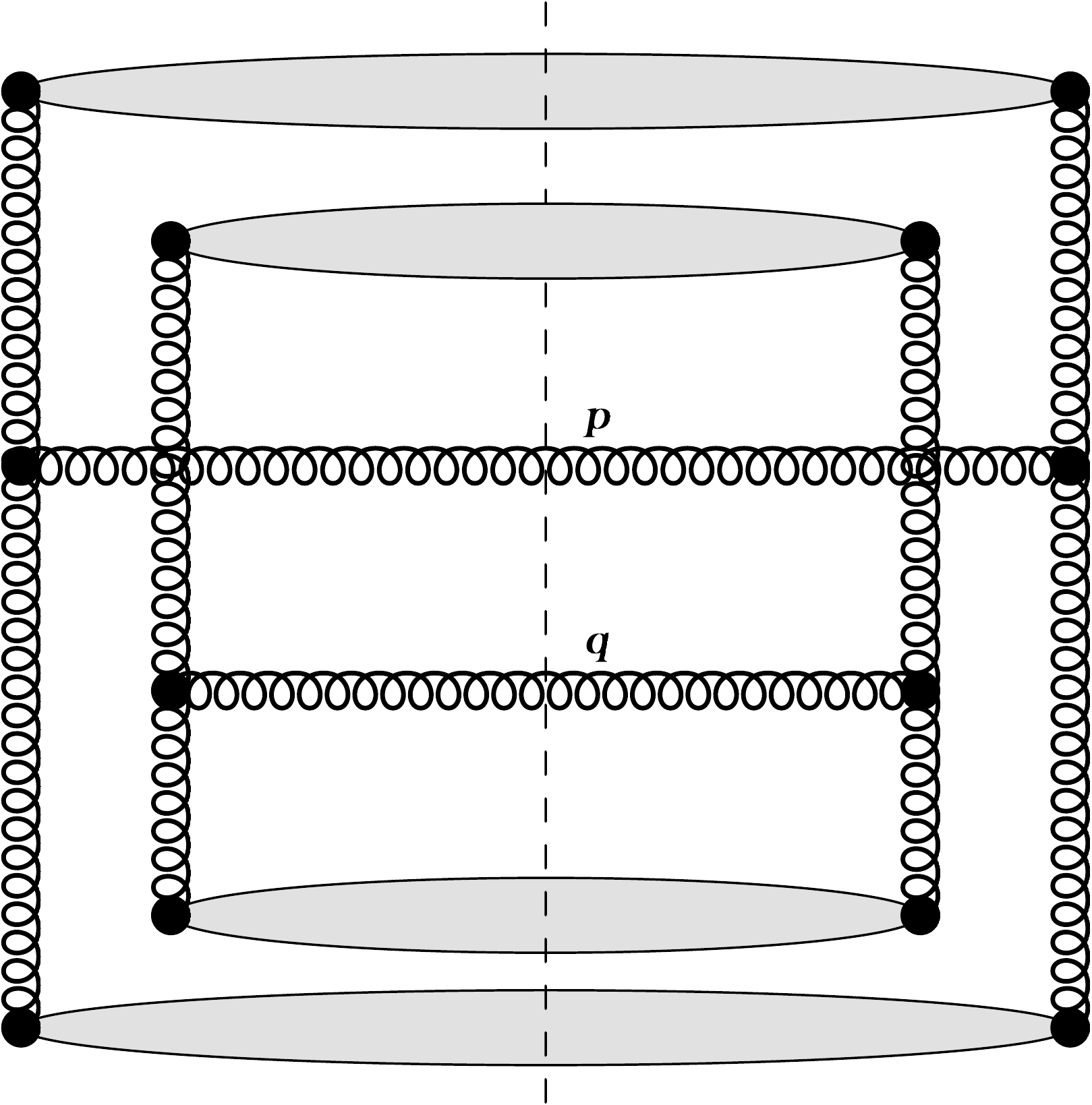}
\caption{Disconnected diagram for inclusive production of two gluons
  with momenta $p$ and $q$.}
\label{fig:SquaredSingle}
\end{figure}
We start with the expression for the two gluon transverse momentum
distribution for glasma graphs given in
ref.~\cite{Altinoluk:2016}\footnote{However, we neglect corrections
  due to the non-zero thickness of the projectile or target derived in
  ref.~\cite{Altinoluk:2016}.}:
\be
\frac{\mathrm{d} N} {\mathrm{d}y_p \mathrm{d}^2p \, \mathrm{d}y_q
  \mathrm{d}^2q} =
16N_{c}^2(N_{c}^2-1) \, g^{12}\, 
\frac{A_{\perp}\Lambda^2}{p^4q^4\Lambda^4}
\frac{\mu_{T}^4\mu_{P}^4}{(2\pi)^2} \, (\mathcal{A}+\mathcal{B}+\mathcal{C})~.
\ee
Here, $\Lambda$ denotes an infrared cutoff for applicability of the
leading twist, weak field approximation. $A_\perp \Lambda^2$ will be taken to be
$\sim 1$ or greater.  $\mathcal{A}$ corresponds to the
disconnected diagram for inclusive double gluon production shown in
fig.~\ref{fig:SquaredSingle};
\begin{figure}[htb]
\includegraphics[scale=.2]{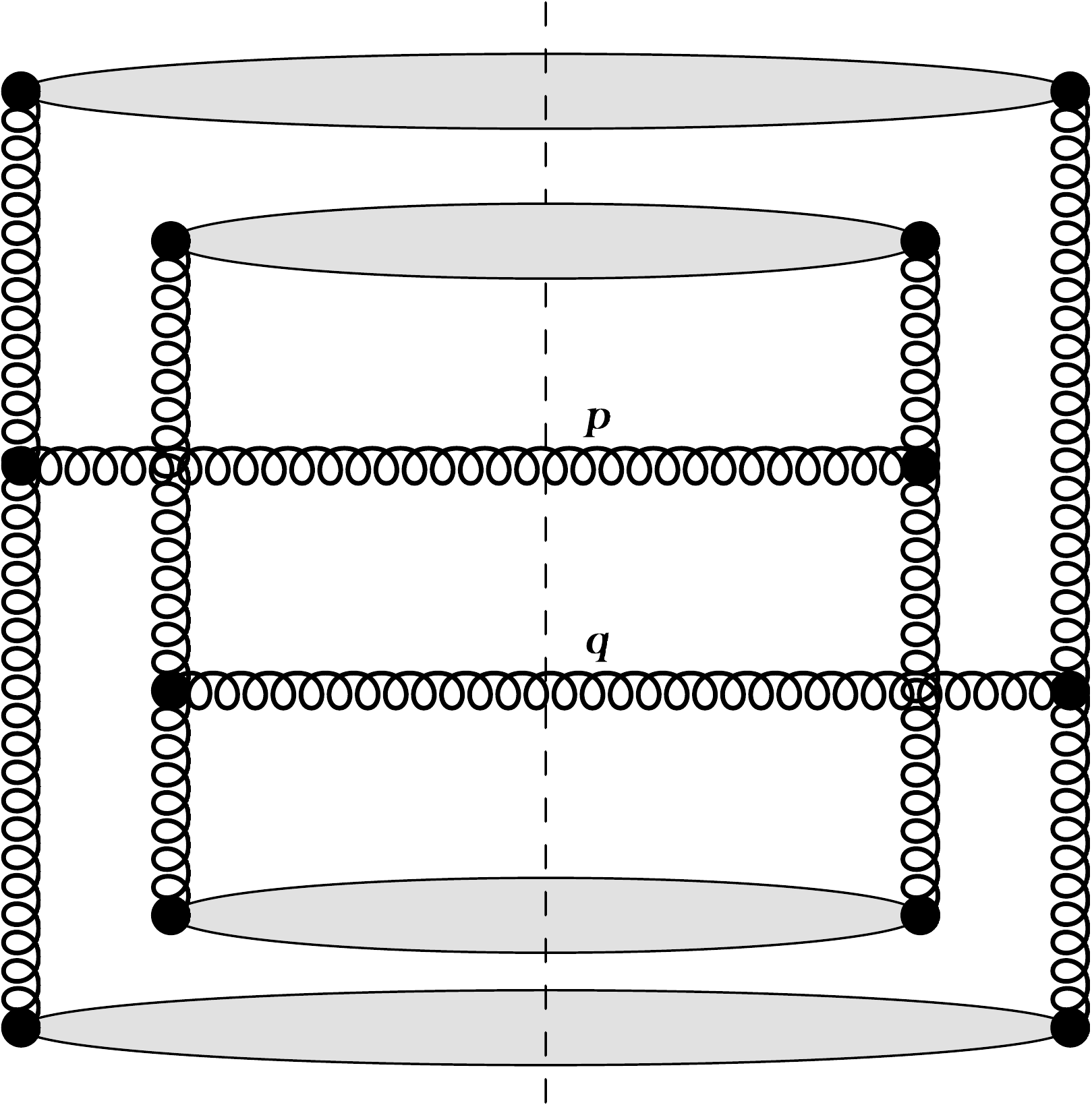}~~~~~~~~~
\includegraphics[scale=.245]{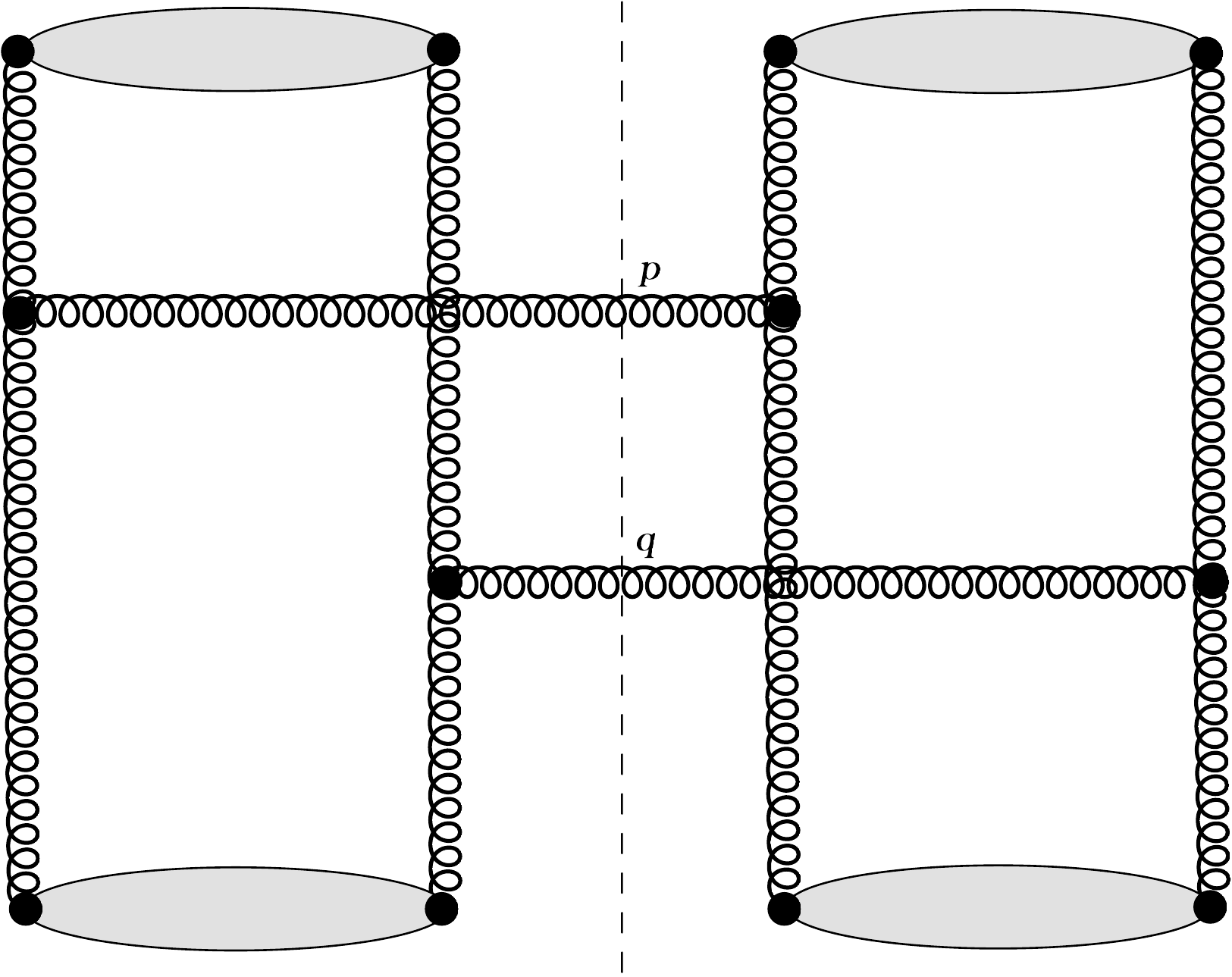}
\caption{HBT diagrams proportional to $\delta(\vec p \pm \vec q)$.}
\label{fig:HBT_diagrams}
\end{figure}
$\mathcal{C}$ are the HBT-like~\cite{Altinoluk:2015eka} parts
proportional to $\delta^2(\vec{p}\pm\vec{q})$, shown in
fig.~\ref{fig:HBT_diagrams}; and the remaining diagrams are combined into
$\mathcal{B}$ (fig.~\ref{fig:B_diagrams}), and have been interpreted as
Bose enhancement~\cite{Altinoluk:2015uaa}. Note that $\mathcal{B}$ and
$\mathcal{C}$ correspond to connected two gluon production diagrams.

Explicitly,
\be
\mu_T^4\mu_P^4g^8\mathcal{A} =
\frac{(N_{c}^2-1)A_{\perp}\Lambda^2p^2q^2}{(2\pi)^2}
\int^\infty_{\Lambda^2} \mathrm{d}^2{k_1}\,
\, \Phi_P(\vec{k_1})\, \Phi_T(\vec{k_1}-\vec{p})
\int^\infty_{\Lambda^2} \mathrm{d}^2{k_2}\,
\, \Phi_P(\vec{k_2})\, \Phi_T(\vec{k_2}-\vec{q})~,
\label{eq:A}
\ee
\begin{dmath}
\mu_T^4\mu_P^4g^8\mathcal{B} = \Lambda^2p^2q^2
\int^\infty_{\Lambda^2} \mathrm{d}^2{k}\,
\, \Phi_T(\vec{k}-\vec{p})\, \Phi_P(\vec{k})
\left[\, \Phi_P(\vec{k})\, \Phi_T(\vec{k}-\vec{q}) +
  \, \Phi_T(\vec{k}-\vec{p})\, \Phi_P(\vec{k}-\vec{p}-\vec{q}) +
  \frac{1}{8}\, \Phi_T(\vec{k}-\vec{p}) \, \Phi_P(\vec{k}-\vec{p}-\vec{q})
  \frac{f(\vec{k},\vec{p},\vec{q})}{(\vec{k}-\vec{p}-\vec{q})^2
    k^2(\vec{k}-\vec{q})^4}\right]
+ {(\vec{q}\rightarrow-\vec{q})}~,
\label{eq:B}
\end{dmath}
where
\begin{dmath}
f(\vec{k}, \vec{p}, \vec{q}) = \left[k^2(\vec{k}-\vec{q})^2+(\vec{p}+
  \vec{q}-\vec{k})^2(\vec{p}-\vec{k})^2-p^2(2\vec{k}-\vec{q}-\vec{p})^2\right]\\
\left[(\vec{p}+\vec{q}-\vec{k})^2(\vec{k}-\vec{q})^2+k^2(\vec{p}-\vec{k})^2-q^2(2\vec{k}-\vec{q}-\vec{p})^2\right]~.
\end{dmath}
The contribution from HBT diagrams is
\begin{dmath}
  \mu_T^4\mu_P^4g^8\mathcal{C} = \frac{\Lambda^2p^2q^2}{4}
  \int^{\infty}_{\Lambda^2}\mathrm{d}^2{k_1}\mathrm{d}^2{k_2}\,
  \Phi_P(\vec{k_1})\, \Phi_P(\vec{k_2})\, \Phi_T(\vec{k_1}-\vec{p})\, \Phi_T(\vec{k_2}-\vec{q})\\
  \left[\delta^2(\vec{p}-\vec{q})
    \left\{1+\frac{k_2^2(\vec{p}-\vec{k_1})^2}{k_1^2(\vec{p}-\vec{k_2})^2}-\frac{p^2(\vec{k_1}-\vec{k_2})^2}{k_1^2(\vec{p}-\vec{k_2})^2}\right\}
    \left\{1+\frac{k_1^2(\vec{q}-\vec{k_2})^2}{k_2^2(\vec{q}-\vec{k_1})^2}-\frac{q^2(\vec{k_1}-\vec{k_2})^2}{k_2^2(\vec{q}-\vec{k_1})^2}\right\}
    +
    \delta^2(\vec{p}+\vec{q})\left\{1+\frac{k_2^2(\vec{p}-\vec{k_1})^2}{k_1^2(\vec{p}+\vec{k_2})^2}-\frac{p^2(\vec{k_1}+\vec{k_2})^2}{k_1^2(\vec{p}+\vec{k_2})^2}\right\}
    \left\{1+\frac{k_1^2(\vec{q}-\vec{k_2})^2}{k_2^2(\vec{q}+\vec{k_1})^2}-\frac{q^2(\vec{k_1}+\vec{k_2})^2}{k_2^2(\vec{q}+\vec{k_1})^2}\right\}\right]~.
  \label{eq:C}
\end{dmath}
Note that contributions ${\cal B}$ and ${\cal C}$, arising from
connected diagrams, come with neither a second power of the transverse
area, $A_\perp$, nor a second factor of $N_c^2-1$. This is because there is a
single connected color flow loop.

It is sufficient for our present purposes to consider a bias on the
target's ensemble of gluon distributions only. We first compute the
contributions denoted by $\mathcal{A}$ and $\mathcal{B}$. We will compute these assuming that
$\vec{p}\neq\pm\vec{q}$, i.e.\ that $\vec{p}\pm\vec{q}$ are hard
momenta themselves, much greater than the saturation scales of the
colliding protons or nuclei. In fact, when the magnitudes of $\vec p$
and $\vec q$ are close, one needs to also account for the
back-to-back dijet contribution (see
e.g.\ ref.~\cite{Dusling:2012iga}) when computing angular correlations
of high-$p_T$ gluons. Therefore, while we do give the expression for
${\cal C}$ for general $\eta(\vec k)$ later in this section for
completeness, we do not consider these contributions further in
sec.~\ref{sec:ModelEnsembles}.

In this paper, we will only consider reflection symmetric gluon
distributions: $\eta(-\vec{p})=\eta(\vec{p})$. To perform the
integrations over the transverse momenta of the gluons in projectile
and target, we expand the integrands in eqs.~(\ref{eq:A}, \ref{eq:B},
\ref{eq:C}) around the singularities of the Coulomb propagators, and
keep the leading terms. For example,
\be
\int\limits_{\Lambda^2} \frac{\mathrm{d}^2{k}}{k^2}
\frac{\eta(\vec{k}-\vec{p})}{(\vec{k}-\vec{p})^2} =
\frac{2\pi}{p^2}\log\frac{p^2}{\Lambda^2}
+
\int\limits_{\Lambda^2} \frac{\mathrm{d}^2{k}}{k^2}
\frac{\eta(\vec{k}-\vec{p})-1}{(\vec{k}-\vec{p})^2}~.  \label{eq:I(p)}
\ee
The first term is the DGLAP logarithm~\cite{DGLAP}. To compute the
integral in the second term, we first write $\eta(\vec k) -1 =
\widetilde\eta(k)\, \Theta(Q^2-k^2)$ to display explicitly the finite
support of the modification to the gluon distribution. Now, if $Q^2$
is on the order of $p^2$, the contribution from small $k^2\ll p^2$
to the integral is
\be  \label{eq:_I1(p)}
I_1(\vec p) = \frac{\pi}{p^2}\, \widetilde\eta(\vec p)\,
\log\frac{p^2}{\Lambda^2}~~~~, ~~~(\text{if}~Q^2 \sim p^2)~.
\ee
This contribution is absent\footnote{To smoothly interpolate from
  $Q^2\sim \Lambda^2$ to $Q^2\sim p^2$, one could replace the logarithm
in eq.~(\ref{eq:_I1(p)}) by $\log p^2/(p^2-Q^2+\Lambda^2)$. However,
we prefer to avoid such ad hoc interpolations, and instead distinguish
small and large $Q^2$ explicitly.} if $Q^2 \ll p^2$. For any $Q^2 \sim p^2$
or less, the integral on the r.h.s.\ of eq.~(\ref{eq:I(p)}) also
receives a contribution from the region $(\vec p-\vec k)^2\ll p^2$,
provided that $\widetilde\eta(\vec\ell)/\ell^2$ has a pole at $\ell\to0$:
\be
I_2(\vec p) = \frac{1}{p^2} \int\limits_{\Lambda^2}^{Q^2}
\frac{\mathrm{d}^2{\ell}}{\ell^2}\,
\widetilde\eta(\vec\ell)~~~~~~~~~,~~~~~~~~~(\text{if}~\widetilde\eta(\vec
\ell)/\ell^2~\text{has a pole at}~\ell\to 0)~.
\ee
Then eq.~(\ref{eq:A}) is approximated as
\be
\mathcal{A} \approx \frac{(N_c^2-1)A_{\perp}\Lambda^2}{(2\pi)^2}
\left[2\pi\log\frac{p^2}{\Lambda^2} + p^2I_1(\vec p) +
  p^2I_2(\vec p)\right]
\left[2\pi\log\frac{q^2}{\Lambda^2} + q^2I_1(\vec q) +
  q^2I_2(\vec q)\right]~.
\label{eq:A_approx}
\ee
Of course, in an unbiased ensemble where $\eta(\vec k)-1 = \widetilde\eta(k)=0$,
this contribution for independent production of two gluons does not
depend on the angle between $\vec p$ and $\vec q$. The same
is true if $\eta(\vec k)$ is isotropic.\\

For the contribution from connected two-gluon production diagrams we obtain
\begin{dmath}
\mathcal{B} \approx \left\{\frac{\Lambda^2}{(\vec{p}+\vec{q})^2}
\left[\frac{q^2}{p^2}
  \int\limits^{\text{min}(p^2,(\vec{p}+\vec{q})^2)}
  \frac{\mathrm{d}^2k}{k^2}\,
  \left[\eta(\vec{k})\eta(\vec{p}+\vec{q})+\eta^2(\vec{p})\right] +
       {(\vec{p}\leftrightarrow\vec{q})}\right] +
\frac{1}{2}\frac{\Lambda^2}{p^2q^2(\vec{p}+\vec{q})^4}
\int\limits^{\text{min}(p^2,q^2,(\vec{p}+\vec{q})^2)}
\frac{\mathrm{d}^2k}{k^4} \, g(\vec{k},\vec{p},\vec{q})\,
\left[\eta^2(\vec{p}) + \eta^2(\vec{q}) + \eta^2(\vec{p}+\vec{q}) +
  \eta^2(\vec{k})\right]\\
+ \frac{\Lambda^2}{2}\int\limits^{\text{min}(p^2,q^2)}\mathrm{d}^2{k}\,
\frac{\eta^2(\vec{k})}{k^4}\right\} +{(\vec{q}\rightarrow-\vec{q})}~.
\label{eq:B_approx}
\end{dmath}
Here,
\be
g(\vec{k},\vec{p},\vec{q})=[p^2\vec{k}\cdot(\vec{p}+\vec{q})-(\vec{p}+\vec{q})^2\vec{k}\cdot\vec{p}]
\,
  [q^2\vec{k}\cdot(\vec{p}+\vec{q})-(\vec{p}+\vec{q})^2\vec{k}\cdot\vec{q}]~,
\ee
is one fourth the leading term of $f(\vec{k},\vec{p},\vec{q})$
in the limit $k^2\ll p^2,q^2$.

The expansion in eq.~(\ref{eq:B_approx}) includes terms that
explicitly depend on the azimuthal angle, $\phi$, between $\vec{p}$
and $\vec{q}$, even though they may be subleading at large $p^2,
q^2$. In contrast, we have dropped a term in eq.~(\ref{eq:B_approx})
that does not depend on $\phi$, and that would be subleading when
$\mathcal{A}$ and $\mathcal{B}$ are added. However, we have not
dropped the last term in eq.~(\ref{eq:B_approx}), which exhibits
power-sensitivity to low transverse momenta but is independent of
$\phi$ when $\eta(\vec k)$ is isotropic. In
sec.~\ref{sec:ModelEnsembles} we will compute the angular moments
\be
\left< e^{in\phi}\right> = \frac{\int
  \frac{\mathrm{d} N}{\mathrm{d}y_p\mathrm{d}^2p\,
    \mathrm{d}y_q\mathrm{d}^2q}\, e^{in\phi}\, \mathrm{d}\phi} {\int
  \frac{\mathrm{d}N}{\mathrm{d}y_p\mathrm{d}^2p\,
    \mathrm{d}y_q\mathrm{d}^2q}\, \mathrm{d}\phi} \rightarrow \frac{\int
  \frac{\mathrm{d}N}{\mathrm{d}y_p\mathrm{d}^2p\,
    \mathrm{d}y_q\mathrm{d}^2q}\, \cos(n\phi)\, \mathrm{d}\phi} {\int
  \frac{\mathrm{d}N}{\mathrm{d}y_p\mathrm{d}^2p\,
    \mathrm{d}y_q\mathrm{d}^2q}\, \mathrm{d}\phi} ~.
\label{eq:ang_avg}
\ee
Reflection symmetry under the simultaneous $\vec p \to - \vec p$,
$\vec q \to - \vec q$ implies invariance under $\phi\rightarrow-\phi$,
and so $\left< e^{in\phi}\right>$ is real.

If $\eta(\vec{k})$ is
isotropic, eq.~(\ref{eq:B_approx}) gives the leading
$\phi$-dependent terms. However, in sec.~\ref{sec:aniso_eta} we shall
see that, when $\eta(\vec{k})$ is anisotropic, eq.~(\ref{eq:A_approx})
will also contribute to the angular moments. In that case, the angular
correlations in the ``disconnected diagrams'' actually arise due to the
bias.

Using the same approximation for the integrations over the 2d Coulomb
propagators, the ``HBT diagrams'' for general $\eta(\vec{k})$ evaluate to
\begin{dmath}
\mathcal{C}\approx \Lambda^2
\left[\delta^2(\vec{p}-\vec{q})+\delta^2(\vec{p}+\vec{q})\right]
\left\{\frac{\pi^2}{2}\eta^2(\vec{p}) \log^2\frac{p^2}{\Lambda^2} +
\int\limits_{\Lambda^2}^{p^2}\frac{\mathrm{d}^2k_1}{k_1^4}
\frac{\mathrm{d}^2k_2}{k_2^4}\eta(\vec{k}_1)
\left[(\vec{k}_1\cdot\vec{k}_2)^2\eta(\vec{k}_2) +
  2\eta(\vec{p}) \left(\vec{k}_1\cdot\vec{k}_2 -
  \frac{2(\vec{k}_1\cdot\vec{p})
    (\vec{k}_2\cdot\vec{p})}{p^2}\right)^2 \right]\right\}~.
\end{dmath}
%

\section{Specific ensembles}
\label{sec:ModelEnsembles}
In this section, we evaluate explicitly the contributions from diagrams
${\cal A}$ and ${\cal B}$ for a few choices of $\eta(\vec k)$. As
already mentioned above, we will focus on the case where the transverse
momenta $\vec p$ and $\vec q$ do not have very similar magnitudes,
and so will ignore diagrams of type ${\cal C}$.

We study a biased ensemble where the number of gluons (defined from
the covariant gauge gluon distribution) with squared transverse
momenta between $\Lambda^2$ and $Q^2$ is boosted. In this section,
we consider ensembles of the form
\be
\eta(\vec{k})=1+\eta_0\,
\frac{\Lambda^{2a}}{k^{2a}}(\hat{k}\cdot\hat{E})^{2b} \,
\Theta(Q^2-k^2)\, \Theta(k^2-\Lambda^2)~, 
\ee
where $\eta_0$ is a dimensionless constant, $a$ controls the
transverse momentum dependence, and $b\ge0$ the anisotropy (in the
direction $\hat{E}$). We consider isotropic ensembles with $a=0$
(section~\ref{sec:biased_ensemble_a=0}) and $a=1$
(section~\ref{sec:biased_ensemble_a=1}), as well as anisotropic
ensembles with $a=0$ and $b=1$ (sec.~\ref{sec:aniso_eta}). We will
also briefly discuss the case $a=-1, b=0$ at the end of
section~\ref{sec:biased_ensemble_a=1}.

\subsection{Unbiased Ensemble} \label{sec:unbiased_ensemble}
Setting $\eta(\vec{k})=1$ in eqs.~(\ref{eq:A_approx},
\ref{eq:B_approx}), we have
\begin{dmath}
\left[\frac{\mathrm{d}N}{\mathrm{d}y_p \mathrm{d}^2p \, \mathrm{d}y_q
  \mathrm{d}^2q}\right]_{\text{unb.}} \approx 16N_{c}^2(N_{c}^2-1) \, g^{12}
\frac{A_{\perp}}{p^4q^4\Lambda^2}
\frac{\mu_{T}^4\mu_{P}^4}{(2\pi)^2}\\
\left\{\frac{1}{2}(N_{c}^2 - 1)(A_{\perp}\Lambda^2)
  \log\frac{p^2}{\Lambda^2}
  \log\frac{q^2}{\Lambda^2}+ \mathcal{O}(1) \\
  +
  2\pi\frac{\Lambda^2}{(\vec{q}+\vec{p})^2}
  \left[\frac{q^2}{p^2} \log
    \frac{\text{min}(p^2, (\vec{q}+\vec{p})^2)}{\Lambda^2}
    +\frac{p^2}{q^2}\log \frac{\text{min}(q^2,
      (\vec{q}+\vec{p})^2)}{\Lambda^2}
    -\frac{1}{2} \left(1+{\vec{q}\cdot\vec{p}} \left( \frac{1}{p^2} +
    \frac{1}{q^2}\right)\right) \log \frac{\text{min}(p^2,q^2,
      (\vec{q}+\vec{p})^2)}{\Lambda^2} \right]\\+
  \mathcal{O}\left(\frac{\Lambda^2}{p^2}\right)\right\} +
          {(\vec{q}\rightarrow-\vec{q})}~.
\label{eq:dN_unbiased}
\end{dmath}
Here, $\mathcal{O}(1)$ stands for the subleading
$\phi$-independent terms, while
$\mathcal{O}\left(\frac{\Lambda^2}{p^2}\right)$ stands for the
subleading $\phi$-dependent terms. Only the $\phi$-dependent terms
enter into the numerator of eq.~(\ref{eq:ang_avg}), from which we can
see that the odd moments vanish. To compute the even moments, we will
need the integrals
\begin{dmath}
  \int\mathrm{d}\phi\cos(2n\phi)\left(\frac{1}{(\vec{p}-\vec{q})^2} +
  \frac{1}{(\vec{p}+\vec{q})^2}\right)
=\frac{4\pi}{|p^2-q^2|}\frac{q_{<}^{2n}}{q_{>}^{2n}}~,
\label{eq:contour}
\end{dmath}
\begin{dmath}
  \int\mathrm{d}\phi\cos(2n\phi) \left(\frac{\vec{q}\cdot\vec{p}}
             {(\vec{p}-\vec{q})^2} - 
  \frac{\vec{q}\cdot\vec{p}}{(\vec{p}+\vec{q})^2}\right) =
  2\pi\frac{p^2+q^2}{|p^2-q^2|} \frac{q_{<}^{2n}}{q_{>}^{2n}}~,
\label{eq:contour1}
\end{dmath}
which can be derived with contour integration. Here
$q_{<}^2=\text{min}(p^2,q^2)$ and $q_{>}^2=\text{max}(p^2,q^2)$.  To
use eqs.~(\ref{eq:contour} , \ref{eq:contour1}) in
(\ref{eq:dN_unbiased}), we need to neglect the dependence of
$(\vec{p}\pm\vec{q})^2$ on $\phi$ when the former appears inside a
logarithm. This is justified in leading logarithmic approximation. The
even angular moments for $n\ge1$ then read
\be \left< e^{2n i
  \phi}\right>_{\text{unb.}} \approx
\frac{\pi}{(N_{c}^2-1)A_{\perp}\Lambda^2} \frac{\Lambda^2}{\mid
  p^2-q^2\mid} \frac{q_{<}^{2n}}{q_{>}^{2n}}\frac{
  \left(\frac{q^2}{p^2}+\frac{p^2}{q^2}\right) \log
  \frac{\text{min}(p^2,q^2,\ell^2)}{\Lambda^2} +
  4\frac{q^2}{p^2}\log\frac{\text{min}(p^2,\ell^2)}{\Lambda^2}+
  4\frac{p^2}{q^2}\log\frac{\text{min}(q^2,\ell^2)}{\Lambda^2}}
     {\log\frac{p^2}{\Lambda^2} \log\frac{q^2}{\Lambda^2}}~,
\label{eq:ang_avg_cross}
\ee
where $\ell^2 = p^2+q^2-2pq$. This formula predicts that $\left< e^{2n
  i \phi}\right>$ decreases with $n$ like $[\text{min}(p^2,q^2) /
  \text{max}(p^2,q^2)]^n$. When~\footnote{As explained in
  sec.~\ref{sec:General_eta}, we do not consider the case where $p$ and
  $q$ are very similar. What we mean here is that their difference
  should be less than $p$ and $q$ themselves (but still much greater
  than the saturation scales of projectile and target).}  $p^2\approx
q^2$, eq.~(\ref{eq:ang_avg_cross}) simplifies to
\be
\left< e^{2n i \phi}\right>_{\text{unb.}} \approx
\frac{10\pi}{(N_{c}^2-1)A_{\perp}\Lambda^2}  \frac{\Lambda^2}{\mid p^2-q^2\mid}\frac{\log\frac{\ell^2}{\Lambda^2}} {\log\frac{p^2}{\Lambda^2}
       \log\frac{q^2}{\Lambda^2}}~.
\ee

\subsection{Constant boost of the gluon density between $k^2 =
  \Lambda^2$ and $k^2 = Q^2$}
\label{sec:biased_ensemble_a=0}

In this section, we consider a boost of the gluon density between $k^2
= \Lambda^2$ and $k^2 = Q^2$ by a constant factor $1+\eta_0$:
\be
\eta(\vec{k})=1+\eta_0\, \Theta(Q^2-k^2)\, \Theta(k^2-\Lambda^2)~.
\label{eq:eta_a=0}
\ee
This corresponds to the ``penalty'' action
\be
V_{\text{eff}} =
\frac{1}{8\pi}\, N_c^2\, A_{\perp} \Lambda^2\, \frac{Q^2-\Lambda^2}{\Lambda^2}\,
     [\eta_0-\log(1+\eta_0)]~,
\label{eq:penalty_a=0}
\ee
and to a gluon number excess
\be
\Delta N_g = \frac{1}{8\pi}\, N_c^2\, A_{\perp} \, g^4\, \mu^2\,
\eta_0\,\log\frac{Q^2}{\Lambda^2}~.
\ee
Hence, for $\eta(k)$ like in eq.~(\ref{eq:eta_a=0}), a substantial
gluon number excess is much more likely to occur due to many
additional gluons with small transverse momenta not too far above
$\Lambda$, so that $\eta_0$ is large but $Q^2/\Lambda^2$ is
moderate. However, such configurations do not increase $\langle
k_T^2\rangle$ much (biasing towards gluon distributions like
  eq.~(\ref{eq:eta_a=0}) with large $\langle k_T^2\rangle$ would
  rather favor smaller $\eta_0$ and larger $Q^2$).

We first calculate the two-gluon spectrum for $Q^2\sim p^2, q^2,
(\vec{p}\pm\vec{q})^2$. Factoring $(1+\eta_0)^2$ from
eqs.~(\ref{eq:A}) and (\ref{eq:B}), we have
\begin{dmath}
\left[\frac{\mathrm{d}N}{\mathrm{d}y_p \mathrm{d}^2p \, \mathrm{d}y_q
  \mathrm{d}^2q}\right]_{\text{bias}}=(1+\eta_0)^2
\left[\frac{\mathrm{d}N}{\mathrm{d}y_p \mathrm{d}^2p \,
    \mathrm{d}y_q
  \mathrm{d}^2q}\right]_{\text{unb.}} ~.
\end{dmath}
We conclude that, in the limit where $Q^2$ is on the order of the
momenta of the produced gluons (or greater), the angular moments in
this ensemble are the same as in the unbiased ensemble,
c.f.\ eq.~(\ref{eq:ang_avg_cross}). This is analogous to the
$k$-independent rescaling of the color charge density of the proton
considered in ref.~\cite{Mace:2018yvl}. Note, however, that such gluon
distributions have very small probability in the original ensemble,
since $V_{\text{eff}} \propto Q^2$.

On the other hand, when $Q^2\ll p^2,q^2,(\vec{p}\pm\vec{q})^2$, we can
drop contributions like those in eq.~(\ref{eq:_I1(p)}).  Then
\begin{dmath}
\left[\frac{\mathrm{d}N}{\mathrm{d}y_p \mathrm{d}^2p \, \mathrm{d}y_q
  \mathrm{d}^2q}\right]_{\text{bias}} \approx 16N_{c}^2(N_{c}^2-1) \, g^{12}
\frac{A_{\perp}}{p^4q^4\Lambda^2}
\frac{\mu_{T}^4\mu_{P}^4}{(2\pi)^2} \\
\left\{\frac{1}{2}(N_{c}^2 - 1)(A_{\perp}\Lambda^2)
  \left(\log\frac{p^2}{\Lambda^2}+ \frac{\eta_0}{2}\log\frac{Q^2}{\Lambda^2}\right)
  \left(\log\frac{q^2}{\Lambda^2}+
  \frac{\eta_0}{2}\log\frac{Q^2}{\Lambda^2} \right)+ \mathcal{O}(1) +
  2\pi\frac{\Lambda^2}{(\vec{q}+\vec{p})^2}
  \left[\frac{q^2}{p^2} \left( \log\frac{\text{min}(p^2,
      (\vec{q}+\vec{p})^2)}{\Lambda^2}
    + \frac{\eta_0}{2}\log\frac{Q^2}{\Lambda^2}\right)
    +\frac{p^2}{q^2}\left( \log \frac{\text{min}(q^2,
      (\vec{q}+\vec{p})^2)}{\Lambda^2}
    + \frac{\eta_0}{2}\log\frac{Q^2}{\Lambda^2}\right)
    -\frac{1}{2} \left( 1+{\vec{q}\cdot\vec{p}}
    \left(\frac{1}{p^2}+\frac{1}{q^2}\right)\right)
    \left(\log\frac{\text{min}(p^2,q^2,(\vec{q}+\vec{p})^2)}{\Lambda^2}
    + \left(\frac{\eta_0}{2}+\frac{\eta_0^2}{4}\right)
    \log\frac{Q^2}{\Lambda^2}\right)\right]\\+
  \mathcal{O}\left(\frac{\Lambda^2}{p^2}\right)\right\}
  +{(\vec{q}\rightarrow-\vec{q})}~.
\label{eq:Xsec_const_eta}
\end{dmath}
It may be surprising, at first glance, that a ``distortion'' of the
gluon distribution up to $Q^2$ much less than the momenta of the
produced gluons would affect the cross section. This is due to the
fact that the production occurs via Lipatov fusion~\cite{bfkl} of one gluon
from the projectile with one gluon from the target, where typically one
of the fusing gluons carries much smaller transverse momentum than the
produced gluon.

The even angular moments for $n\ge1$ resulting from
eq.~(\ref{eq:Xsec_const_eta}) are
\begin{dmath}
\langle e^{2n i \phi}\rangle_{\text{bias}} \approx
\frac{\pi}{(N_{c}^2-1)A_{\perp}\Lambda^2}  \frac{\Lambda^2}{\mid p^2-q^2\mid}
\frac{q_{<}^{2n}}{q_{>}^{2n}}\\
\frac{
\left(\frac{q^2}{p^2}+\frac{p^2}{q^2}\right)\left(
\log\frac{\text{min}(p^2,q^2,\ell^2)}{\Lambda^2}
+ \left(\frac{5\eta_0}{2}+\frac{\eta_0^2}{4}\right)
\log\frac{Q^2}{\Lambda^2}\right)
+4\frac{q^2}{p^2}\log\frac{\text{min}(p^2,\ell^2)}{\Lambda^2}
+4\frac{p^2}{q^2}\log\frac{\text{min}(q^2,\ell^2)}{\Lambda^2}}
     {\left(\log\frac{p^2}{\Lambda^2}+ \frac{\eta_0}{2}\log\frac{Q^2}{\Lambda^2}\right)
  \left(\log\frac{q^2}{\Lambda^2}+ \frac{\eta_0}{2}\log\frac{Q^2}{\Lambda^2} \right)}~.
\label{eq:dipole_a=0}
\end{dmath}
We recall that $\ell^2 = p^2+q^2-2pq$.

We have yet to make any assumptions about the magnitude of
$\eta_0$. First, consider $\eta_0=\mathcal{O}(1)$, with $Q^2$ less than
$\text{min}(p^2,q^2,\ell^2)$. This corresponds to a class of high
gluon multiplicity configurations with $\langle k_T^2\rangle$
moderately higher than in the absence of the bias. Here, the
corrections to the numerator and denominator of
eq.~(\ref{eq:dipole_a=0}) due to the bias are suppressed only
logarithmically (relative to the unbiased ensemble)~!  On the other
hand, if $\eta_0\ll1$, we can simplify the previous expression by
ignoring the terms $\sim\eta_0^2$ to compute the ratio of the moments
in the biased and unbiased ensembles\footnote{Note that the
  dependence on $n$ cancels in these ratios.}. For $\ell^2<
\text{min}(p^2,q^2)$, for example, this is
\be
\frac{\langle e^{2 n i \phi}\rangle_{\text{bias}}} {\langle e^{2 n i
    \phi}\rangle_{\text{unb.}}}
\approx 1 + \frac{\eta_0}{2} \frac{\log\frac{Q^2}{\Lambda^2}}
        {\log\frac{\ell^2}{\Lambda^2}}
        \left(1-\frac{\log\frac{\ell^2}{\Lambda^2}}
             {\log\frac{q^2}{\Lambda^2}}-
             \frac{\log\frac{\ell^2}{\Lambda^2}}
                  {\log\frac{p^2}{\Lambda^2}}\right) ~.
\label{eq:dipole_ratio}
\ee
Thus, when $\vec p$ and $\vec q$ have comparable magnitudes ($\to
\ell^2 \ll p^2, q^2$), the correction is positive, and so the
angular moments in the biased ensemble increase with increasing
$\eta_0$. In the limit $p^2\gg q^2$ (or vice versa), on the other
hand, the correction in eq.~(\ref{eq:dipole_a=0}) is negative,
\be
\frac{\langle e^{2 n i \phi}\rangle_{\text{bias}}} {\langle e^{2 n i
    \phi}\rangle_{\text{unb.}}}
\approx 1-\frac{\eta_0}{2}\frac{\log\frac{Q^2}{\Lambda^2}}
        {\log\frac{p^2}{\Lambda^2}}~,
\label{eq:dipole_ratio1}
\ee
and so the angular moments decrease with increasing $\eta_0$.  Thus,
we see that even a very simple ``distortion'' of the gluon
distribution may give rise to a fairly intricate behavior of the
angular correlations.
%

\subsection{Transverse momentum dependent boost of the gluon density}
\label{sec:biased_ensemble_a=1}

We now consider a transverse momentum dependent increase of the gluon
density between $k^2 = \Lambda^2$ and $k^2=Q^2$. We will mostly
confine ourselves to an $\eta(k)$ that decreases with increasing $k$, 
but will briefly take up the case of $\eta(k)$ increasing with $k$
at the end of this section. Let's then start with
\be
\eta(k)=1 + \eta_0\,\frac{\Lambda^{2}}{k^{2}}\,
\Theta(Q^2-k^2)\, \Theta(k^2-\Lambda^2)~,
\label{eq:eta_a}
\ee
with $Q^2 \gg \Lambda^2$.  Such a gluon distribution comes with a
penalty action of
\be
V_{\text{eff}}[\eta] = \frac{1}{8\pi}\,N_c^2 \, A_{\perp} \Lambda^2 \,
\left[
  \left(\frac{Q^2}{\Lambda^2}-\eta_0\right) \log \left(1+\eta_0
  \frac{\Lambda^2}{Q^2}\right) + (1+\eta_0)\log(1+\eta_0)\right]~.
\ee
Also, the number of high-$k_T$ gluons in the hadron increases by
\be
\Delta N_g[\eta] = \frac{1}{8\pi}\,N_c^2 \, A_{\perp}\,
g^4\, \mu^2\, \eta_0~.
\ee
Hence, both $V_{\text{eff}}$ and $\Delta N_g$ approach a constant as
$Q^2/\Lambda^2 \gg 1$ at fixed $\eta_0$.

The resulting gluon production distribution is
\begin{dmath}
\left[\frac{\mathrm{d}N}{\mathrm{d}y_p \mathrm{d}^2p \, \mathrm{d}y_q
  \mathrm{d}^2q}\right]_{\text{bias}} \approx 16N_{c}^2(N_{c}^2-1) \, g^{12}
\frac{A_{\perp}}{p^4q^4\Lambda^2}
\frac{\mu_{T}^4\mu_{P}^4}{(2\pi)^2}\\
\left\{\frac{1}{2}(N_{c}^2 - 1)(A_{\perp}\Lambda^2)
  \left(\log\frac{p^2}{\Lambda^2}+ \frac{\eta_0}{2}\right)
  \left(\log\frac{q^2}{\Lambda^2}+ \frac{\eta_0}{2}\right)+\mathcal{O}(1)
  + \frac{\pi}{3}\eta_0^2 +
  2\pi\frac{\Lambda^2}{(\vec{q}+\vec{p})^2}
  \left[\frac{q^2}{p^2}
    \left( \log \frac{\text{min}(p^2, (\vec{q}+\vec{p})^2)}{\Lambda^2}+
    \frac{\eta_0}{2}\right)+\frac{p^2}{q^2}
    \left( \log \frac{\text{min}(q^2, (\vec{q}+\vec{p})^2)}{\Lambda^2}+
    \frac{\eta_0}{2}\right) -
    \frac{1}{2} \left(1+{\vec{q}\cdot\vec{p}}
    \left(\frac{1}{p^2}+\frac{1}{q^2}\right)\right)
    \left(\log\frac{\text{min}(p^2,q^2, (\vec{q}+\vec{p})^2)}{\Lambda^2}+
    \frac{\eta_0}{2}+\frac{\eta_0^2}{8}\right)\right]+
  \mathcal{O}\left(\frac{\Lambda^2}{p^2}\right)\right\}
+{(\vec{q}\rightarrow-\vec{q})}~.
\end{dmath}
No logarithms of $Q^2$ appear here, since the gluon distribution in
this biased ensemble, eq.~(\ref{eq:eta_a}), drops more rapidly than $1/k^4$.
The angular moments are
\begin{dmath}
\langle e^{2n i \phi}\rangle_{\text{bias}} \approx
\frac{\pi}{(N_{c}^2-1)A_{\perp}\Lambda^2}  \frac{\Lambda^2}{\mid p^2-q^2\mid}
\frac{q_{<}^{2n}}{q_{>}^{2n}} \\
~~~~\frac{
\left(\frac{q^2}{p^2}+\frac{p^2}{q^2}\right)\left(
\log\frac{\text{min}(p^2,q^2,\ell^2)}{\Lambda^2}+
\frac{5\eta_0}{2}+\frac{\eta_0^2}{8}\right) +
4\frac{q^2}{p^2}\log\frac{\text{min}(p^2,\ell^2)}{\Lambda^2}+
4\frac{p^2}{q^2}\log\frac{\text{min}(q^2,\ell^2)}{\Lambda^2}}
     {\left(\log\frac{p^2}{\Lambda^2}+\frac{\eta_0}{2}\right)
       \left(\log\frac{q^2}{\Lambda^2}+\frac{\eta_0}{2}\right)+
       \frac{2\pi\eta_0^2}{3\Lambda^2A_{\perp}(N_c^2-1)}}~.
\label{eq:a=1}
\end{dmath}
To simplify this expression, we consider various parametric
magnitudes for $\eta_0$. The case $\eta_0\lsim 1$ is not very
interesting, as it does not lead to rare configurations with a
substantial increase in the number of gluons; we have
$V_{\text{eff}}\propto N_c^2\, A_\perp \Lambda^2 \, \eta_0$ and
$\Delta N_g\propto N_c^2 \, A_{\perp}\, g^4\, \mu^2\, \eta_0$,
independent of $Q^2$.

It is more interesting to let $\eta_0\sim \sqrt{\log
  \frac{Q^2}{\Lambda^2}}$, so that the amplitude of the shift of the
gluon distribution and its transverse momentum cutoff are
related. $\Delta N_g$ then increases in proportion to
$\sqrt{\log\frac{Q^2}{\Lambda^2}}$, while $V_{\text{eff}} \propto
\sqrt{\log\frac{Q^2}{\Lambda^2}}\,
\log\left(\log\frac{Q^2}{\Lambda^2}\right)$.  This means that high
gluon multiplicities can be reached with much higher probability than
for the bias considered in the previous section.  Furthermore, the
angular moments will increase with $\eta_0^2$ for any choice of $p^2$
or $q^2$. For example, when $p^2\approx q^2$,
\be
\frac{\langle e^{2 n i \phi}\rangle_{\text{bias}}} {\langle e^{2 n i
    \phi}\rangle_{\text{unb.}}}
\approx 1+\frac{1}{40}\frac{\eta_0^2}{\log\frac{\ell^2}{\Lambda^2}}~,
\label{eq:limit_a=1}
\ee
and the correction increases like the square of $\Delta N_g$,
i.e.\ proportional to $\log Q^2$.

We briefly comment on the case where the gluon distribution at the
saddle point of the reweighted ensemble is shifted to
\be
\eta(k)=1 + \eta_0\,\frac{k^{2}}{\Lambda^{2}}\,
\Theta(Q^2-k^2)\, \Theta(k^2-\Lambda^2)~,
\label{eq:eta_a=-1}
\ee
again with $Q^2\gg \Lambda^2$.  This leads to a very strong increase
of $\langle k_T^2\rangle$ as compared to the unbiased ensemble. On the
other hand, to have $V_{\text{eff}} \propto {Q^2}$ like in
sec.~\ref{sec:biased_ensemble_a=0}, rather than $V_{\text{eff}} \propto
{Q^4}$, we must choose very small amplitude, $\eta_0\propto
1/Q^{2}$. And then, $\Delta N_g$ asymptotes to a constant at large
$Q^2$. In other words, these are rare configurations of the hadron
where the excess mean squared transverse momentum of the gluons is
large but their number excess is not. In this paper we restrict to
cases corresponding to large $\Delta N_g$.

\subsection{Anisotropic $\eta(\vec{k})$}
\label{sec:aniso_eta}
In this section, we explore anisotropic gluon distributions such that
the average gluon distribution $X_s(k)$ in the unbiased ensemble
is multiplied by
\be \label{eq:eta(K)_anisotropic}
\eta(\vec{k})=1+\eta_0\frac{\Lambda^{2a}}{k^{2a}}(\hat{k}\cdot\hat{E})^{2b}
\, \Theta(Q^2-k^2)\, \Theta(k^2-\Lambda^2)~.
\ee
The vector $\hat E$ specifies an arbitrary direction in the transverse
plane.  Studying such anisotropic configurations has been suggested by
Kovner and Lublinsky~\cite{Kovner:2011pe} (also see
ref.~\cite{Dumitru:2014vka}). In their work, the anisotropy is due to
fluctuations from configuration to configuration of the hadron or
nucleus. In this scenario, after computing the angular correlator,
$\langle e^{i n \phi}\rangle$, we would have to perform an average over
the directions of $\hat E$. In what follows, we will instead consider the
possibility that the 2d rotational symmetry is explicitly broken due to
an external bias (like a spin model in an external magnetic field), so
that eq.~(\ref{eq:eta(K)_anisotropic}) represents the gluon
distribution {\em averaged over the reweighted ensemble}. It is beyond
the scope of the present paper to discuss specific phenomenological
models for how such an anisotropic bias may arise in p+p or p+A
collisions. Nevertheless, it is an interesting exercise to compute
two-particle correlations in an anisotropic ensemble, as this leads to
new contributions to the angular moments.

Specifically, the disconnected diagram, $\mathcal{A}$, will now
contribute $\phi$-dependent terms. From
eqs.~(\ref{eq:_I1(p)},~\ref{eq:A_approx}), we see that these terms only
occur when $Q^2\sim p^2,q^2$, and that they will be proportional to
$\eta_0^2\, \left[(\hat{p}\cdot\hat{E})^{2}\,
  (\hat{q}\cdot\hat{E})^{2}\right]^b$. Therefore, only moments less
than or equal to $2b$ will receive such contributions.

When $a=1$, the leading angular contribution to $\mathcal{A}$ is
proportional to $\frac{\Lambda^4}{p^4}\log^2\frac{p^2}{\Lambda^2}$,
which is smaller than the angular contributions we saw in
sec~\ref{sec:biased_ensemble_a=0}. Hence, we consider the case $a=0$
and $b=1$. We then have, for $\eta_0\ll1$, $Q^2\gg\Lambda^2$,
\be
V_{\text{eff}} \approx \frac{3}{128\pi}\, N_c^2\, A_{\perp}
\, Q^2\, \eta_0^2~,
\label{eq:Veff_aniso}
\ee
and
\be
\Delta N_g= \frac{1}{16\pi}\, N_c^2\, A_{\perp} \, g^4\, \mu^2\,
\eta_0\,\log\frac{Q^2}{\Lambda^2}~.
\ee

The contribution from the disconnected diagram now becomes
(for $Q^2\sim p^2,q^2$)
\be
\mathcal{A}_{\text{bias}} \approx (N_c^2-1)A_{\perp}\Lambda^2 \log
\frac{p^2}{\Lambda^2} \log \frac{q^2}{\Lambda^2} \left[
  1+\frac{\eta_0}{4}+
  \frac{\eta_0^2}{2}(\hat{p}\cdot\hat{E})^2 \right]\left[1+\frac{\eta_0}{4}+ \frac{\eta_0^2}{2}(\hat{q}\cdot\hat{E})^2
\right]~.
\label{eq:A_aniso}
\ee
Because this is the contribution from the ``disconnected diagram'',
and there is a constant $k$-independent boost of the gluon
distribution relative to the unbiased ensemble, we recover the two
DGLAP logarithms even in the anisotropic contribution.

The anisotropic part of $\mathcal{B}$ at linear order in $\eta_0$
is\footnote{We do not include the contribution to $\mathcal{B}$ of
  order $\eta_0^2$ because it is a subleading contribution to the part
  of the cross section of order $\eta_0^2$, and because we restrict to
  the elliptic anisotropy in this section.}
\begin{dmath}
\mathcal{B}_{\text{bias}}-\mathcal{B}_{\text{unb.}}\approx\frac{\pi\eta_0}{4}\frac{\Lambda^2}{(\vec{p}+\vec{q})^2}\left\{\left(8(\hat{p}\cdot\hat{E})^2+4[\hat{E}\cdot(\widehat{p+q})]^2+2\right)\left[\frac{q^2}{p^2}\log\frac{\text{min}(p^2,(\vec{p}+\vec{q})^2)}{\Lambda^2}-\frac{1}{8}\left(1+\vec{q}\cdot\vec{p}\left(\frac{1}{p^2}+\frac{1}{q^2}\right)\right)\log\frac{\text{min}(p^2,q^2,(\vec{p}+\vec{q})^2)}{\Lambda^2}\right]+\log\frac{\text{min}(p^2,q^2,(\vec{p}+\vec{q})^2)}{\Lambda^2}\left[(\hat{p}\cdot\hat{q})^2[\hat{E}\cdot(\widehat{p+q})]^2+\frac{\hat{E}\cdot\vec{p}}{p^2}\hat{E}\cdot\left(\vec{q}-\frac{\vec{p}\cdot\vec{q}}{p^2}(\vec{p}+\vec{q})\right)\right]+{(\vec{p}\leftrightarrow\vec{q})}\right\}+{(\vec{q}\rightarrow-\vec{q})}~.
\label{eq:B_aniso}
\end{dmath}
(vectors with a hat denote unit vectors).  Eqs.~(\ref{eq:A_aniso},
\ref{eq:B_aniso}) both give rise to non-zero odd angular moments.
Integrating $\mathcal{A}_{\text{bias}}\, \cos n\phi$ over $\phi$, we get
\be
\int\mathrm{d}\phi\cos(n\phi)\mathcal{A}_{\text{bias}} \approx
(N_c^2-1) A_{\perp} \Lambda^2
\log\frac{p^2}{\Lambda^2} \log\frac{q^2}{\Lambda^2}\, \frac{\pi}{32}\,
\begin{cases}
(16\eta_0+8\eta_0^2)\cos(\psi-2\phi_E) & n=1\\
\eta_0^2 & n=2\\
0 & n\ge2
\end{cases} ~~~,
\ee
where $\psi$ is the ``center of mass angle" (the average of the angles
made by $\vec{p}$ and $\vec{q}$) and $\phi_E$ is the angle made by
$\hat{E}$. If the anisotropy is due to fluctuations, and we average
over the direction $\phi_E$ of $\hat E$, then only the contribution to
the $n=2$ elliptic
moment is non-zero. On the other hand, for an external bias with fixed
direction, there is a non-zero $n=1$ moment when the average azimuthal
angle of $\vec p$ and $\vec q$ is not equal to that of $\hat E$ plus
$45^\circ$.

In the following, we focus on the elliptical
anisotropy. By comparing eq.~(\ref{eq:A_aniso}) with
eq.~(\ref{eq:B_aniso}), we see that, if we choose $1\gg \eta_0 \gg
\frac{\Lambda^2}{Q^2} \left(\log\frac{Q^2}{\Lambda^2}\right)^{-1}$,
then the main contribution to the angular moments will be from
$\mathcal{A}_{\text{bias}}$, with $\mathcal{B}_{\text{bias}}$ a small
correction. The elliptical anisotropy is then
\begin{dmath}
\langle e^{2 i \phi}\rangle_{\text{bias}} \approx\langle e^{2 i
  \phi}\rangle_{\text{unb.}}+\frac{\eta_0^2}{64}~,
\label{eq:aniso_avg}
\end{dmath}
which increases with increasing $\eta_0$. The remarkable aspect of
this expression is that the correction due to the bias is independent
of $p$ and $q$, and that it will dominate the azimuthal correlation
for sufficiently large $p^2$, $q^2$, $|p^2-q^2|$.

\section{Summary and Conclusion}

Imposing a bias on the functional integral over small-$x$ gluon
distributions in a hadron (or photon or nucleus) modifies the
expectation value of the gluon distribution, as well as statistical
fluctuations about it. One may bias with respect to the multiplicity
of gluons, their average squared transverse momentum, their transverse
energy, or, more generally, towards a specific modification of their
distribution over transverse momentum. Another example is p+p
collision events with a hard jet or a $Z$-boson. Each of these biases
produces a {\em distinct} ``distortion'' of the average gluon
distribution.  The modification of various observables under such
biases provides, in principle, a test of our understanding of
high-energy QCD, a fact well known to and appreciated by developers of
``event generators'' for high-energy collisions. Our specific focus
here (in continuation of prior work in
refs.~\cite{Dumitru:2017cwt,Dumitru:2018iko}) is on relating the bias
to reweighting of the functional integral over the effective action
for small-$x$ gluons.\\

In the present paper, we analyzed the effect of some simple model
biases on the two-gluon transverse momentum spectrum, as well as their
azimuthal correlations, in a high-energy collision.  We found
interesting differences between the biased ensembles that were
studied. For example, when the bias boosts the gluon density
uniformly by a constant factor, for all $k$ up to the hard momenta $p$ and $q$ of the produced
gluons, there is no effect on the azimuthal
moments. Such a modification, however, has essentially infinite
action, and probability zero.  On the other hand, if one increases
the density of gluons uniformaly by a constant factor, but only over transverse momentum squared $\ll p^2,
q^2, (\vec p \pm \vec q)^2$, the azimuthal moments {\em are} affected
by the bias. This is due to the fact that the Lipatov process involves
fusion of two gluons, one with $k$ much less than $p, q$ and the other
with $k$ comparable to $p, q$. We find that the elliptic angular
moment, $\langle \exp(2i\phi)\rangle$, increases with the gluon number excess
$\Delta N_g$ when $\vec{p}$ and $\vec{q}$ have comparable magnitude,
but decreases with $\Delta N_g$ when $p\gg q$.

Another example of at least theoretical interest is that of a bias
that introduces a preferred direction in the transverse impact
parameter plane, inducing an anisotropy of the gluon distribution.
Such a bias gives rise to an angular dependence even of the
disconnected (in terms of color flow) diagram for double gluon
production. Hence, the leading dependence of the cross section on the
azimuthal angle $\phi$ between $\vec p$ and $\vec q$ emerges at
leading power in $N_c$, and comes with the same DGLAP logarithms $\log
p^2\, \log q^2$ as the diagrams for uncorrelated production in an
unbiased ensemble.

Our analysis could be continued by looking at more complicated biases
and how they would reweight the functional integral over gluon
distributions. In particular, a very important step would be to
develop a more direct connection between the reweighting functional
$b[X]$ used here and specific event ensembles that one might be able
to construct in practice, either from experiment or from an event
generator.

\section*{Acknowledgements}

I gratefully acknowledge support by the DOE Office of Nuclear Physics
through Grant No.\ DE-FG02-09ER41620; and from The City University of
New York through the PSC-CUNY Research grant 62098-0050.  I thank
A.~Dumitru for numerous useful discussions and suggestions and for
help with the preparation of the manuscript. I also thank V.~Skokov
for comments on the draft.

\end{document}